\documentclass[10pt]{article}
\usepackage{graphicx}
\begin{document}
\title{The antiJaynes-Cummings model is solvable : quantum Rabi model in rotating and counter-rotating frames ; following the experiments}
\author{Joseph Akeyo Omolo\\
Department of Physics, Maseno University, P.O. Private Bag, Maseno, Kenya\\
e-mail:~ ojakeyo04@yahoo.co.uk ~;~ ojakeyo@maseno.ac.ke}
\date{20 February 2021}
\maketitle

\begin{abstract}
This article is a response to the continued assumption, cited even in reports and reviews of recent experimental breakthroughs and advances in theoretical methods, that the antiJaynes-Cummings (AJC) interaction is an intractable energy non-conserving component of the quantum Rabi model (QRM). We present three key features of QRM dynamics : (a) the AJC interaction component has a conserved excitation number operator and is exactly solvable (b) QRM dynamical space consists of a rotating frame (RF) dominated by an exactly solved Jaynes-Cummings (JC) interaction specified by a conserved JC excitation number operator which generates the $U(1)$ symmetry of RF and a correlated counter-rotating frame (CRF) dominated by an exactly solved antiJaynes-Cummings (AJC) interaction specified by a conserved AJC excitation number operator which generates the $U(1)$ symmetry of CRF (c) for QRM dynamical evolution in RF, the initial atom-field state $\vert e0\rangle$ is an eigenstate of the effective AJC Hamiltonian $\overline{H}_{AJC}$, while the effective JC Hamiltonian $H_{JC}$ drives this initial state $\vert e0\rangle$ into a time evolving entangled state, and, in a corresponding process for QRM dynamical evolution in CRF, the initial atom-field state $\vert g0\rangle$ is an eigenstate of the effective JC Hamiltonian, while the effective AJC Hamiltonian drives this initial state $\vert g0\rangle$ into a time evolving entangled state, thus addressing one of the long-standing challenges of theoretical and experimental QRM dynamics; consistent generalizations of the initial states $\vert e0\rangle$ ,
$\vert g0\rangle$ to corresponding $n\ge0$ entangled eigenstates $\vert~\overline\Psi_{en}^{~+}\rangle$ ~,~ $\vert\Psi_{gn}^-\rangle$ of the AJC in RF and JC in CRF, respectively, provides general dynamical evolution of QRM characterized by collapses and revivals in the time evolution of the atomic, field mode, JC and AJC excitation numbers for large initial photon numbers ; the JC and AJC excitation numbers are conserved in the respective frames RF , CRF, but each evolves with time in the alternate frame.

\end{abstract}

\section{Introduction}
The quantum Rabi model (QRM) is the simplest form of quantized light-matter interaction coupling a single two-level atom to a single mode of quantized light. The Hamiltonian of the system takes the standard form
$$H_R=\hbar\omega\left(\hat{a}^\dagger\hat{a}+\frac{1}{2}\right)+\hbar\omega_0s_z+\hbar g(\hat{a}+\hat{a}^\dagger)(s_-+s_+) \ ; \quad\quad s_z=\frac{1}{2}\sigma_z \ ; \quad \sigma_x=s_-+s_+ \eqno(1)$$
where $(\hat{a} , \hat{a}^\dagger , \omega)$ and $(s_z , s_- , s_+ , \omega_0)$ are the respective quantized field mode and atomic spin state operators and frequencies with standard meanings, while $g$ is the coupling constant. Opening the brackets in the interaction term reveals that the Hamiltonian $H_R$ is composed of a rotating component, the Jaynes-Cummings (JC) interaction and a counter-rotating component, the antiJaynes-Cummings (AJC) interaction, which we obtain explicitly below. Since the two components are algebraically correlated in the sense that they do not commute, the quantum Rabi Hamiltonian $H_R$ must be considered in its full form, which has made it too difficult, if not impossible, to determine the exact general time evolving state of QRM.

Great advances have been made in developing fairly accurate theoretical methods [1-5] to gain insight into QRM dynamics beyond the rotating wave approximation (RWA). However, a major drawback has remained the long-standing assumption that the antiJaynes-Cummings component does not have a conserved excitation number and is not exactly solvable, even though it is now several years since the present author constructed the AJC excitation number operator and proved its conservation in [6], followed by simpler reformulation and exact solutions of both JC and AJC dynamics in [7 , 8]. The good news is that experiments on the full QRM have now made outstanding breakthroughs, determining details of the dynamics someway into the USC-DSC regimes [9-12]. However, in both theoretical and experimental advances, the lack of information on the solvability of the AJC interaction has meant that the coupling range is defined only from the JC side where the rotating wave approximation (RWA) applies, considering the USC-DSC regimes far beyond the RWA as the dynamical region dominated by the counter-rotating component of QRM. Since the main purpose of the present article is to supply the long-missing information on the conserved excitation number operator and exact solutions of the AJC interaction, we do not give details of the general historical and technical developments of the outstanding theoretical and experimental achievements made so far on QRM, but refer the interested readers to the most recent excellent reviews [13 , 14 , 15].

This article is organized as follows : we present the basic algebraic and dynamical structure of QRM in section~$2$, where the JC and AJC components are defined together with the respective conserved excitation number, $U(1)$ and parity operators, as well as the rotating and counter-rotating frames RF , CRF, leading to the corresponding rotating and counter-rotating wave approximations RWA , CRWA ; we develop dynamics in RF (RWA) and CRF (CRWA) in section~$3$, where we provide consistent generalizations of the initial states $\vert e0\rangle$ , $\vert g0\rangle$ to the corresponding $n\ge0$ entangled eigenstates of the AJC and JC in RF , CRF, respectively ; the Conclusion is given in section~$4$.

\section{Basic algebraic and dynamical structure of QRM}
An important point to note is that the algebraic properties of the basic state operators of a fully quantized system such as QRM directly determine the dynamical structure of the system. To achieve a clear understanding of the internal structure of QRM dynamics, we begin by applying the algebraic properties of the atom-field state operators $\hat{a}$ , $\hat{a}^\dagger$ , $s_z$ , $s_-$ , $s_+$ to develop the basic algebraic structure,  automatically leading us to identification of two algebraically correlated dynamical frames of QRM as we now present in this section.

\subsection{Basic algebraic structure of QRM}
Following the original work in [6] where details are presented, we apply normal and antinormal operator ordering of the basic atomic and quantized field mode state operators to express the standard quantum Rabi model (QRM) Hamiltonian $H_R$ in equation (1) in the symmetrized rotating and counter-rotating component form [6-8]
$$H_R=\frac{1}{2}(H+\overline{H}~) \eqno(2a)$$
The rotating component $H$ is the Jaynes-Cummings (JC) Hamiltonian obtained in the normal order form
$$H=\hbar\omega\hat{N}+\hbar\delta s_z+2\hbar g(\hat{a}s_++\hat{a}^\dagger s_-)-\frac{1}{2}\hbar\omega \quad ; \quad \hat{N}=\hat{a}^\dagger\hat{a}+s_+s_- \quad ; \quad \delta=\omega_0-\omega \eqno(2b)$$
where $\hat{N}$ is the standard conserved excitation number operator of the JC interaction, commuting with the Hamiltonian $H$ according to
$$[\hat{N}~,~H]=0 \eqno(2c)$$
which we have proved explicitly in [6-8] and is easily proved here using the JC qubit state transition operator introduced in equation (5$c$). The parameter $\delta=\omega_0-\omega$ in equation (2$b$) is the usual JC red-sideband frequency detuning.

The counter-rotating component $\overline{H}$ is the antiJaynes-Cummings (AJC) Hamiltonian obtained in the antinormal order form
$$\overline{H}=\hbar\omega\hat{\overline{N}}+\hbar\overline{\delta}s_z+2\hbar g(\hat{a}s_-+\hat{a}^\dagger s_+)-\frac{1}{2}\hbar\omega \quad ; \quad \hat{\overline{N}}=\hat{a}\hat{a}^\dagger+s_-s_+ \quad ; \quad \overline{\delta}=\omega_0+\omega \eqno(2d)$$
where $\hat{\overline{N}}$ is the (long-missing) conserved excitation number operator of the AJC interaction, which we first constructed and proved conserved in [6], commuting with the Hamiltonian $\overline{H}$ according to
$$[~\hat{\overline{N}}~,~\overline{H}~]=0 \eqno(2e)$$
This commutation relation is easily proved here using the AJC qubit state transition operator introduced in equation (7$b$) below. The parameter $\overline{\delta}=\omega_0+\omega$ in equation (2$d$) is the AJC blue-sideband frequency detuning.

The JC and AJC excitation number operators $\hat{N}$ , $\hat{\overline{N}}$ commute, but each does not commute with the alternate Hamiltonian according to
$$[~\hat{N}~,~\hat{\overline{N}}~]=0 \ ; \quad\quad  [~\hat{N}~,~\overline{H}~]\ne0 \ ; \quad\quad [~\hat{\overline{N}}~,~H~]\ne0 \ ; \quad\quad [~H~,~\overline{H}~]\ne0 \eqno(2f)$$
where the last commutation relation shows that the JC and AJC components do not commute and are therefore algebraically correlated as stated earlier.

With the specification of the basic algebraic structure provided above, we now present the $U(1)$ and parity symmetry properties of QRM generated by the JC and AJC excitation number operators $\hat{N}$ , $\hat{\overline{N}}$.

\subsection{QRM in the rotating frame : RF}
The free evolution operator $U_0(t)=e^{-i\omega t\hat{N}}$ generated by the JC excitation number operator $\hat{N}$ is a $U(1)$ symmetry operator which effects transformation of the JC and AJC Hamiltonians in equations (2$b$) , (2$d$) to the rotating frame (RF) according to [6]
$$U_0(t)=e^{-i\omega t\hat{N}}~:\quad U_0^\dagger(t)HU_0(t)=H \ ; \quad U_0^\dagger(t)~\overline{H}~U_0(t)=\hbar\omega(~\hat{\overline{N}}-\frac{1}{2})+\hbar\overline{\delta}s_z +
2\hbar g(e^{-2i\omega t}\hat{a}s_-+e^{2i\omega t}\hat{a}^\dagger s_+) \eqno(3a)$$
Using equations (2$a$) , (2$b$) , (3$a$), we obtain the transformation of the QRM Hamiltonian $H_R$ to the rotating frame reorganized in the final form
$$U_0^\dagger(t)H_RU_0(t)=H_{JC}+\hbar g(e^{-2i\omega t}\hat{a}s_-+e^{2i\omega t}\hat{a}^\dagger s_+) \quad ; \quad H_{JC}=\hbar\omega\hat{N}+\hbar\delta s_z+\hbar g(\hat{a}s_++\hat{a}^\dagger s_-) \eqno(3b)$$
where $H_{JC}$ is the effective JC Hamiltonian in the rotating frame.

According to equation (3$b$), the effective JC Hamiltonian $H_{JC}$ dominates over the fast oscillating AJC interaction component in the RF. Hence, QRM dynamics in the RF is dominated by the JC interaction mechanism characterized by red-sideband transitions generated by the effective JC Hamiltonian $H_{JC}$. Dropping the fast oscillating AJC component in equation (3$b$), we obtain the QRM Hamiltonian in a rotating wave approximation (RWA) in the RF in the form
$$U_0^\dagger H_RU_0(t)\approx H_{JC} \eqno(3c)$$

\subsection{QRM in the counter-rotating frame : CRF}
The free evolution operator $\overline{U}_0(t)=e^{-i\omega t\hat{\overline{N}}}$ generated by the AJC excitation number operator $\hat{\overline{N}}$ is a $U(1)$ symmetry operator which effects transformation of the JC and AJC Hamiltonians in equations (2$b$) , (2$d$) to the counter-rotating frame (CRF) according to [6]
$$\overline{U}_0(t)=e^{-i\omega t\hat{\overline{N}}}~:\quad \overline{U}_0^\dagger(t)~\overline{H}~\overline{U}_0(t)=\overline{H} \ ; \quad \overline{U}_0^\dagger(t)~H~\overline{U}_0(t)=\hbar\omega(\hat{N}-\frac{1}{2})+\hbar\delta s_z +
2\hbar g(e^{-2i\omega t}\hat{a}s_++e^{2i\omega t}\hat{a}^\dagger s_-) \eqno(3d)$$
Using equations (2$a$) , (2$d$) , (3$d$), we obtain the transformation of the QRM Hamiltonian $H_R$ to the counter-rotating frame reorganized in the final form
$$\overline{U}_0^\dagger(t)~H_R~\overline{U}_0(t)=\overline{H}_{AJC}+\hbar g(e^{-2i\omega t}\hat{a}s_++e^{2i\omega t}\hat{a}^\dagger s_-) \quad ; \quad \overline{H}_{AJC}=\hbar\omega(~\hat{\overline{N}}-1)+\hbar\overline{\delta}s_z+\hbar g(\hat{a}s_-+\hat{a}^\dagger s_+) \eqno(3e)$$
where $\overline{H}_{AJC}$ is the effective AJC Hamiltonian in the counter-rotating frame.

According to equation (3$e$), the effective AJC Hamiltonian $\overline{H}_{AJC}$ dominates over the fast oscillating JC interaction component in the CRF. Hence, QRM dynamics in the CRF is dominated by the AJC interaction mechanism characterized by blue-sideband transitions generated by the effective AJC Hamiltonian $\overline{H}_{AJC}$. Dropping the fast oscillating JC component in equation (3$e$), we obtain the QRM Hamiltonian in a counter-rotating wave approximation (CRWA) in the CRF in the form
$$\overline{U}_0^\dagger~\overline{H}_R~\overline{U}_0(t)\approx \overline{H}_{AJC} \eqno(3f)$$

\subsection{QRM parity symmetry}
An important algebraic property has arisen that QRM is characterized by two commuting excitation number operators, namely, the JC conserved excitation number operator $\hat{N}$ which generates $U(1)$ symmetry in the rotating frame according to equations (3$a$)-(3$c$), and the AJC conserved excitation number operator $\hat{\overline{N}}$ which generates $U(1)$ symmetry in the counter-rotating frame according to equations (3$d$)-(3$f$). The two excitation number operators commute with the respective RF/CRF Hamiltonians $H_{JC}$ , $\overline{H}_{AJC}$, but each does not commute with the Hamiltonian in the alternate frame according to
$$[~\hat{N}~,~H_{JC}~]=0 \ ; \quad [~\hat{N}~,~\overline{H}_{AJC}~]\ne0 \ ; \quad [~\hat{\overline{N}}~,~\overline{H}_{AJC}~]=0 \ ; \quad [~\hat{\overline{N}}~,~H_{JC}~]\ne0 \ ; \quad [~H_{JC}~,~\overline{H}_{AJC}~]\ne0 \eqno(4a)$$
where we recall $[~\hat{N}~,~\hat{\overline{N}}~]=0$ in equation (2$f$). We observe that equation (4$a$) essentially re-emphasizes the commutation relations in equation (2$f$), now within the dynamical frames RF , CRF.

The $U(1)$ symmetry generated by either excitation number operator $\hat{N}$ or $\hat{\overline{N}}$ in the respective frame $RF$ or $CRF$ is therefore not a common symmetry of the full QRM. However, in [6], we established that setting $\omega t=k\pi$, $k=1 , 2 , 3 , ...$, in $U_0(t)$ , $\overline{U}_0(t)$ in equations (3$a$) , (3$d$), provides the parity operator $\hat{\Pi}_k=(~\hat{\Pi}~)^k$ which generates parity symmetry as a common symmetry of QRM according to [6]
$$\hat{\Pi}_k=e^{-ik\pi\hat{N}}=e^{-ik\pi\hat{\overline{N}}}=(~\hat{\Pi}~)^k \quad\ ; \quad\ \hat{\Pi}=e^{-i\pi\hat{N}}=e^{-i\pi\hat{\overline{N}}} \quad ; \quad k=1 , 2 , 3 , ... \eqno(4b)$$
$$\hat{\Pi}_k^\dagger~H~\hat{\Pi}_k=H \quad ; \quad \hat{\Pi}_k^\dagger~\overline{H}~\hat{\Pi}_k=\overline{H} \quad\Rightarrow\quad \hat{\Pi}_k^\dagger~H_R~\hat{\Pi}_k=H_R \eqno(4c)$$
where $\hat{\Pi}$ in equation (4$b$) is the standard QRM parity operator in usual definition with respect to the JC excitation number operator $\hat{N}$, but now given equivalent definition in terms of the AJC excitation number operator $\hat{\overline{N}}$. The equality arising in the corresponding definition with respect to the AJC excitation number operator $\hat{\overline{N}}$ is easily proved [6], using the relation (recall $\hat{a}\hat{a}^\dagger=\hat{a}^\dagger\hat{a}+1 ~,~ s_+s_-+s_-s_+=1$)
$$\hat{\overline{N}}=\hat{N}+2s_-s_+~:\quad e^{-i\pi\hat{\overline{N}}}=e^{-i\pi\hat{N}}e^{-2i\pi s_-s_+} \quad ; \quad e^{-2i\pi s_-s_+}=I \quad\Rightarrow\quad e^{-i\pi\hat{\overline{N}}}=e^{-i\pi\hat{N}}=\hat{\Pi} \eqno(4d)$$
This parity symmetry has been applied to obtain approximate solutions of QRM in the deep strong coupling (DSC) regime [2] and exact stationary state solutions in [3], noting also the related dynamical evolution in [1].

\section{QRM dynamics in RF and CRF : following experiments}
It follows from subsections~$2.2-2.3$ that QRM has two dynamical frames, namely, the rotating frame (RF) and the counter-rotating frame (CRF). Dynamics in RF is dominated by the JC interaction mechanism and through a $U(1)$ symmetry transformation generated by the conserved JC excitation number operator $\hat{N}$, the QRM Hamiltonian is approximated by an effective JC Hamiltonian $H_{JC}$ in a rotating wave approximation (RWA) according to equations (3$b$)-(3$c$). Dynamics in CRF is dominated by the AJC interaction mechanism and through a $U(1)$ symmetry transformation generated by the conserved AJC excitation number operator $\hat{\overline{N}}$, the QRM Hamiltonian is approximated by an effective AJC Hamiltonian $\overline{H}_{AJC}$ in a counter-rotating wave approximation (CRWA) according to equations (3$e$)-(3$f$). This specification of the QRM dynamical frames which we have achieved here has never been done earlier, since CRF was never known to have a conserved excitation number operator, until the present author constructed the correct form $\hat{\overline{N}}$ and proved conservation in [6]. By a comparison with the theoretical models [1 , 2 , 4 , 13 , 14] and experimental designs [5 , 9-12 , 14 , 15] of the full QRM dynamics, RF as specified in the present article may be identified with the JC (weak-strong coupling) regime, but CRF as specified here may not be interpreted as the USC-DSC regime where the RWA breaks down and the counter-rotating terms (AJC interaction) start contributing to the dynamics. We note that the definition of CRF in subsection~$2.3$ does not involve the dimensionless coupling parameter $\frac{g}{\omega}$, which is explicitly used in characterizing the USC-DSC regimes [2 , 4 , 13 , 14], meaning that CRF, even though dominated by the effective AJC interaction, may not necessarily be equivalent to the USC-DSC regime. The property established earlier in [6] and in the present article that QRM dynamical frames RF and CRF are each specified by a corresponding conserved excitation number operator $\hat{N}$ , $\hat{\overline{N}}$, respectively, means that these commuting excitation number operators may be used as order parameters for characterizing QRM dynamics in RF and CRF regions of the general quantum state space. According to the commutation relations in equation (4$a$), the JC excitation number operator $\hat{N}$ is conserved in the dynamics generated by the effective JC Hamiltonian $H_{JC}$ in RF, but non-conserved in the dynamics generated by the effective AJC Hamiltonian $\overline{H}_{AJC}$ in CRF, while the AJC excitation number operator $\hat{\overline{N}}$ is conserved in the dynamics generated by the effective AJC Hamiltonian $\overline{H}_{AJC}$ in CRF, but non-conserved in the dynamics generated by the effective JC Hamiltonian $H_{JC}$ in RF. The dynamical property that the JC excitation number operator $\hat{N}$ is conserved in the RF (RWA) coupling regime, but evolves with time in the USC-DSC regime has been determined in QRM experiments [9]. We now follow the initial states specified in the experiments and apply the RF and CRF specifications in subsections~$2.2-2.3$ to determine and clarify the observed physical features of QRM dynamics. The experiments have focussed attention on QRM dynamics evolving from an initial state with the field mode in the vacuum state and the atom in excited or ground state specified by $\vert e0\rangle$ or $\vert g0\rangle$ in standard notation. In the present work, we provide consistent generalizations of these atom-field initial states to include field mode initial states $\vert n\ge0\rangle$.

\subsection{QRM dynamics in RF}
We have established in subsection~$2.2$ that QRM dynamics in RF is generated by the effective JC Hamiltonian $H_{JC}$ in RWA according to equations (3$b$)-(3$c$). The experiments [9-12] have identified the state $\vert e0\rangle$ with the field mode in the vacuum state $\vert 0\rangle$ and the atom in the excited state $\vert e\rangle$ as the appropriate initial state for QRM dynamics in RF (JC regime). Here, we establish that the initial state $\vert e0\rangle$ in RF is an eigenstate of the effective AJC Hamiltonian $\overline{H}_{AJC}$ and provide a consistent generalization to the corresponding $n\ge0$ initial AJC eigenstate which reduces to $\vert e0\rangle$ for $n=0$.

\subsubsection{Dynamics from initial state $\vert e0\rangle$}
We introduce appropriate notation $\vert\psi_{e0}\rangle$ for the initial state and $\vert\psi_{g1}\rangle$ for the associated transition state defined in standard notation as
$$\vert\psi_{e0}\rangle=\vert e0\rangle \quad\quad ; \quad\quad \vert\psi_{g1}\rangle=\vert g1\rangle \eqno(5a)$$
We observe that the effective AJC Hamiltonian $\overline{H}_{AJC}$ does not generate dynamical evolution of the initial state $\vert e0\rangle$ in RF, since $\vert e0\rangle$ is an eigenstate of $\overline{H}_{AJC}$, which, using $\overline{H}_{AJC}$ from equation (3$e$) and applying standard atom and field mode state algebraic operations, is easily established to satisfy an eigenvalue equation (recall $\overline{\delta}=\omega_0+\omega$)
$$\overline{H}_{AJC}\vert\psi_{e0}\rangle=\frac{1}{2}\hbar(\omega_0+\omega)\vert\psi_{e0}\rangle \eqno(5b)$$
where we identify the energy eigenvalue $\frac{1}{2}\hbar(\omega_0+\omega)$ as the atomic excited state energy $\frac{1}{2}\hbar\omega_0$ and the field mode vacuum state energy $\frac{1}{2}\hbar\omega$ as expected. Equation (5$b$) means that the effective AJC Hamiltonian $\overline{H}_{AJC}$ generates only plane wave evolution $e^{-\frac{it}{2}(\omega_0+\omega)}\vert\psi_{e0}\rangle$ of the initial state $\vert\psi_{e0}\rangle$ which does not describe the general QRM dynamics in RF. However, in agreement with the theoretical models and experiments, the effective JC Hamiltonian $H_{JC}$ generates dynamical evolution of the initial state $\vert\psi_{e0}\rangle$ into a time evolving entangled state which describes the general QRM dynamics in RF as we now demonstrate.

We introduce a JC qubit state transition operator $\hat{A}$ and apply standard algebraic properties of the atom and field mode state operators $\hat{a}$ , $\hat{a}^\dagger$ , $s_z$ , $s_-$ , $s_+$ to determine the relation with the JC excitation number operator $\hat{N}$ in the form
$$\hat{A}=\delta s_z+g(\hat{a}s_++\hat{a}^\dagger s_-) \quad\quad ; \quad\quad \hat{A}^2=\frac{1}{4}\delta^2+g^2\hat{N} \eqno(5c)$$
Using $\hat{A}$ from equation (5$c$), we express the effective JC Hamiltonian $H_{JC}$ defined in equation (3$b$) in the form
$$H_{JC}=\hbar\omega\hat{N}+\hbar\hat{A} \eqno(5d)$$
Conservation of the JC excitation number in the dynamics generated by $H_{JC}$ in RF is easily proved by using the relations in equations (5$c$) , (5$d$) to show that $\hat{N}$ commutes with $H_{JC}$, thus confirming equations (2$f$) , (4$a$) and simplifying the earlier proof in [6].

Applying $\hat{A}$ from equation (5$c$) on the initial state $\vert\psi_{e0}\rangle$ defined in equation (5$a$), using standard atom-field state algebraic operations and reorganizing as appropriate, we obtain JC qubit states $\vert\psi_{e0}\rangle$ , $\vert\phi_{e0}\rangle$ satisfying qubit state transition algebraic operations in the form
$$\hat{A}~\vert\psi_{e0}\rangle=R_{e0}\vert\phi_{e0}\rangle \ ; \quad\quad \hat{A}~\vert\phi_{e0}\rangle=R_{e0}\vert\psi_{e0}\rangle \eqno(5e)$$
where $\vert\phi_{e0}\rangle$ is an entangled qubit transition state obtained in the form
$$\vert\phi_{e0}\rangle=c_{e0}\vert\psi_{e0}\rangle+s_{e0}\vert\psi_{g1}\rangle \ ; \quad R_{e0}=g\sqrt{1+\xi^2} \ ; \quad c_{e0}=\frac{\delta}{2R_{e0}} \ ; \quad s_{e0}=\frac{g}{R_{e0}} \ ; \quad \xi=\frac{\delta}{2g} \eqno(5f)$$
We identify $R_{e0}$ as the Rabi frequency of the JC qubit oscillations. Repeated application of $\hat{A}$ on $\vert\psi_{e0}\rangle$ even and odd number times using equation (5$e$) gives useful general qubit state transition algebraic operations
$$\hat{A}^{2k}~\vert\psi_{e0}\rangle=R_{e0}^{2k}~\vert\psi_{e0}\rangle \quad\quad ; \quad\quad \hat{A}^{2k+1}~\vert\psi_{e0}\rangle=R_{e0}^{2k+1}~\vert\phi_{e0}\rangle \quad ; \quad k=0 , 1 , 2 , 3 , ... \eqno(5g)$$

The general time evolving state $\vert\Psi_{e0}(t)\rangle$ describing QRM dynamics generated by $H_{JC}$ from the initial state $\vert\psi_{e0}\rangle$ is obtained in the form
$$\vert\Psi_{e0}(t)\rangle=U_{JC}(t)\vert\psi_{e0}\rangle \quad\quad ; \quad\quad U_{JC}(t)=e^{-\frac{it}{\hbar}H_{JC}} \eqno(6a)$$
where $U_{JC}(t)$ is the JC time evolution operator which on substituting $H_{JC}$ from equation (5$d$) and noting the commutation relation $[~\hat{N}~,~\hat{A}~]=0$ takes the factorized form
$$U_{JC}(t)=e^{-it\hat{A}}e^{-i\omega t\hat{N}} \eqno(6b)$$
Substituting this into equation (6$a$) and applying $\hat{N}=\hat{a}^\dagger\hat{a}+s_+s_-$ on $\vert\psi_{e0}\rangle$ gives
$$\hat{N}\vert\psi_{e0}\rangle=\vert\psi_{e0}\rangle \quad\quad\Rightarrow\quad\quad
\vert\Psi_{e0}(t)\rangle=e^{-i\omega t}e^{-it\hat{A}}\vert\psi_{e0}\rangle \eqno(6c)$$
Expanding $e^{-it\hat{A}}$ and reorganizing in even and odd power terms
$$e^{-it\hat{A}}=\sum_{k=0}^\infty\frac{(-1)^kt^{2k}}{(2k)!}\hat{A}^{2k}-i\sum_{k=0}^\infty\frac{(-1)^kt^{2k+1}}{(2k+1)!}\hat{A}^{2k+1} \eqno(6d)$$
and substituting into equation (6$c$), applying the general qubit state algebraic operations from equation (5$g$), then introducing standard trigonometric expansions, we obtain the general time evolving state in the final form
$$\vert\Psi_{e0}(t)\rangle=e^{-i\omega t}(\cos(R_{e0}t)\vert\psi_{e0}\rangle-i\sin(R_{e0}t)\vert\phi_{e0}\rangle) \eqno(6e)$$
which describes Rabi oscillations at frequency $R_{e0}$ between the initial separable state $\vert\psi_{e0}\rangle$ and an entangled transition state $\vert\phi_{e0}\rangle$. Substituting $\vert\phi_{e0}\rangle$ from equation (5$f$) into equation (6$e$), reorganizing and introducing the definitions of $\vert\psi_{e0}\rangle$ , $\vert\psi_{g1}\rangle$ from equation (5$a$) reveals that in general, the time evolving state $\vert\Psi_{e0}(t)\rangle$ is a normalized entangled state obtained in the form
$$\vert\Psi_{e0}(t)\rangle=e^{-i\omega t}(~(\cos(R_{e0}t)-ic_{e0}\sin(R_{e0}t)~)\vert e0\rangle-is_{e0}\sin(R_{e0}t)\vert g1\rangle) \ ; \quad\quad \langle\Psi_{e0}(t)\vert\Psi_{e0}(t)\rangle=1 \eqno(6f)$$
Hence, as we set out to demonstrate, the effective JC Hamiltonian $H_{JC}$ generates dynamical evolution of the initial atom-field state
$\vert e0\rangle$ into a time evolving entangled state $\vert\Psi_{e0}(t)\rangle$ in RF. We can now compare QRM dynamics described by $\vert\Psi_{e0}(t)\rangle$ in RF to the corresponding dynamical features observed in the JC ($\frac{g}{\omega}=0.04$) regime in the QRM simulation experiment in [9]. In this respect, we determine the atomic excitation number, field mode mean photon number, the JC and AJC excitation numbers in the state $\vert\Psi_{e0}(t)\rangle$.

Applying the JC excitation number operator $\hat{N}$ on $\vert\Psi_{e0}(t)\rangle$ gives an eigenvalue equation
$$\hat{N}\vert\Psi_{e0}(t)\rangle=\vert\Psi_{e0}(t)\rangle \quad\quad\Rightarrow\quad\quad \overline{N}(t)=1 \eqno(6g)$$
from which it follows that the JC excitation number $\overline{N}(t)$ is conserved in the QRM dynamics generated by the effective JC Hamiltonian $H_{JC}$ in RF as expected from the corresponding commutation relation $[~\hat{N}~,~H_{JC}~]=0$ in equation (4$a$). The experiment in [9] confirmed conservation of the JC excitation number (the only total excitation number known at the time of the experiment) in the corresponding JC regime characterized by $\frac{g}{\omega}=0.04$ in [9].

We determine the atomic population inversion $\overline{s}_z(t)$ and excitation number $\overline{s_+s_-}(t)$, the field mode mean photon number $\overline{n}(t)$ and the AJC excitation number $\overline{\overline{N}}(t)$ in the QRM time evolving state $\vert\Psi_{e0}(t)\rangle$ in RF in the form (recall $s_+s_-=\frac{1}{2}+s_z ~;~ s_-s_+=\frac{1}{2}-s_z$)
$$\overline{s}_z(t)=\frac{1}{2}(1-2s_{e0}^2\sin^2(R_{e0}t)) \ ; \quad\ \overline{s_+s_-}(t)=\frac{1}{2}+\overline{s}_z(t) \ ; \quad\ \overline{n}(t)=s_{e0}^2\sin^2(R_{e0}t) \ ; \quad\ \overline{\overline{N}}(t)=2(1-\overline{s}_z(t)) \eqno(6h)$$
It follows immediately from equation (6$h$) that the AJC excitation number $\overline{\overline{N}}(t)$ is non-conserved and evolves in time in the QRM dynamics generated by the effective JC Hamiltonian $H_{JC}$ in RF as expected from the corresponding commutation relation $[~\hat{\overline{N}}~,~H_{JC}~]\ne0$ in equation (4$a$). This dynamical evolution was not investigated in [9], since the authors were not yet aware of the existence of the AJC excitation number operator as a QRM order parameter at the time of the experiment in [9].

The form of dynamical evolution of the atomic excitation number $X=\overline{s_+s_-}(t)$ and the field mode mean photon number $F=\overline{n}(t)$, characterized by pure Rabi oscillations in Fig.$1$ , Fig.$2$, respectively, agrees precisely with the corresponding dynamical evolution observed in the JC ($\frac{g}{\omega}=0.04$) regime in the experiment in [9]. Fig.$3$ reveals the conservation of the JC excitation number $\overline{N}(t)$ in RF, agreeing with the behavior observed in the JC regime in [9]. The dynamical evolution of the AJC excitation number $\overline{\overline{N}}(t)$ in Fig.$4$ was not investigated in [9] as we have already explained.

\begin{figure}[ph]
\centering
\includegraphics[width=0.30\linewidth]{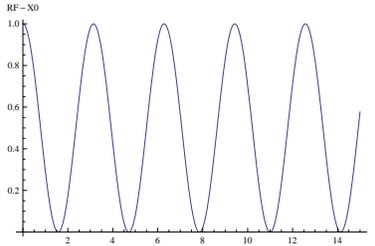}
\caption{JC-atomic excitation number in RF $\overline{s_+s_-}(\tau)~,~\tau=gt :\quad \xi=0~;~\varepsilon=...$}
\label{Fig}
\end{figure}

\begin{figure}[ph]
\centering
\includegraphics[width=0.30\linewidth]{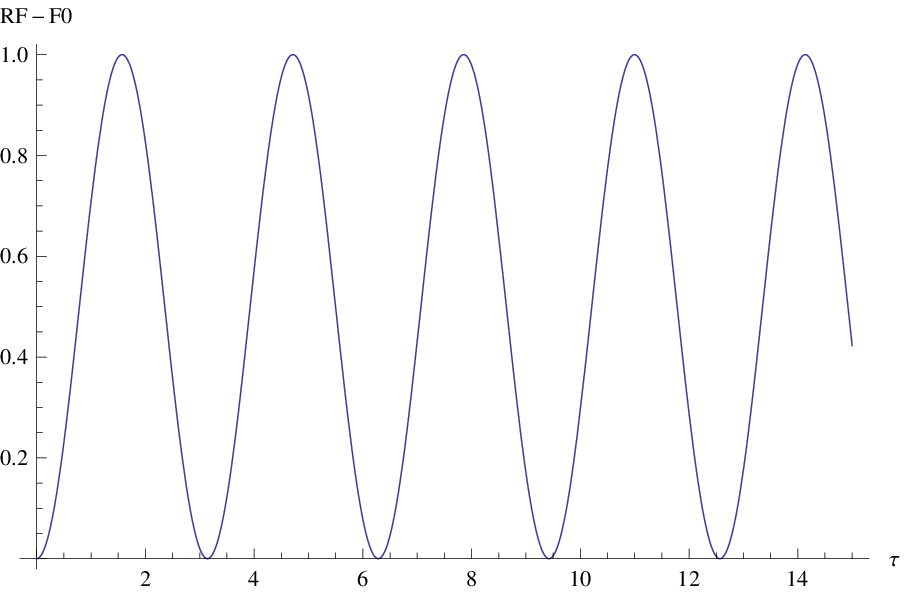}
\caption{JC-field mode mean photon number in RF $\overline{n}(\tau)~,~\tau=gt :\quad \xi=0~;~\varepsilon=...$}
\label{Fig}
\end{figure}

\begin{figure}[ph]
\centering
\includegraphics[width=0.30\linewidth]{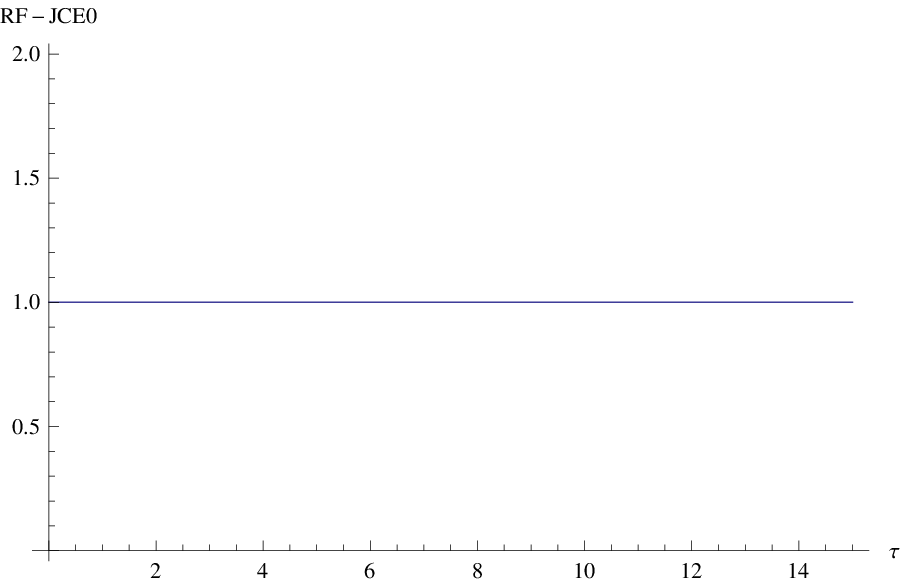}
\caption{JC-excitation number in RF $\overline{N}(\tau)~,~\tau=gt :\quad \xi=0~;~\varepsilon=...$}
\label{Fig}
\end{figure}

\newpage

\begin{figure}[ph]
\centering
\includegraphics[width=0.30\linewidth]{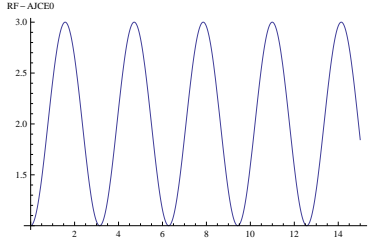}
\caption{AJC-excitation number in RF $\overline{\overline{N}}(\tau)~,~\tau=gt :\quad \xi=0~;~\varepsilon=...$}
\label{Fig}
\end{figure}
There are two points to note at this stage : (i) in presenting the dynamical evolution in Fig.$1$-Fig.$4$, we have set the dimensionless coupling parameter $\xi=\frac{\delta}{2g}=0$ coinciding with $\omega_0=\omega$ in [9], but the parameter $\varepsilon^{-1}=\frac{g}{\omega}$ used to characterize the coupling regimes in [9] does not affect the dynamical evolution described by $\vert\Psi_{e0}(t)\rangle$ in RF, noting $\delta=0$, $\xi=0$ gives $R_{e0}=g$, $c_{e0}=0$, $s_{e0}=1$, $\vert\Psi_{e0}(t)\rangle=e^{-i\omega t}(~\cos(R_{e0}t)\vert e0\rangle-i\sin(R_{e0}t)\vert g1\rangle)$ according to equations (5$f$) , (6$f$) (ii) taking note of the dynamical features of the USC-DSC regimes regarding the vacuum state [1 , 4 , 5 , 11 , 12], the atom-field initial state in [9] is defined with the field mode in a displaced vacuum state with displacement parameter $\pm\frac{g}{\omega}$, while the atom may be in an eigenstate of the spin operator $\sigma_x=s_-+s_+$ so that the general dynamical evolution of the atom-field initial state takes the form of a Schroedinger cat state, which depends on the coupling regime parameter $\frac{g}{\omega}$. Varying $\frac{g}{\omega}$ from the JC to the USC-DSC regimes then determines the dynamical features of QRM as observed in the three coupling regimes in [9]. To obtain corresponding dynamical features in the description developed in the present article, we provide a consistent generalization of the atom-field initial state $\vert e0\rangle$ to an $n\ge0$ entangled eigenstate of the effective AJC Hamiltonian $\overline{H}_{AJC}$ here below.

%\newpage
\subsubsection{Dynamics from a general initial AJC eigenstate}
Noting that the initial state $\vert\psi_{e0}\rangle$ used above in developing QRM dynamics in RF is an eigenstate of the AJC Hamiltonian $\overline{H}_{AJC}$ according to equation (5$b$), we now provide a consistent generalization to an $n\ge0$ initial AJC eigenstate. Since the atom starts in the excited state $\vert e\rangle$, the basic $n\ge0$ atom-field state is $\vert en\rangle$. Considering that the state algebraic operation for determining a general eigenstate of the AJC Hamiltonian couples the state $\vert en\rangle$ to the state $\vert gn-1\rangle$, we introduce appropriate notation $\vert\psi_{en}\rangle$ , $\vert\psi_{gn-1}\rangle$ for the two states in the form
$$\vert\psi_{en}\rangle=\vert en\rangle \ ; \quad\quad \vert\psi_{gn-1}\rangle=\vert gn-1\rangle \eqno(7a)$$
Introducing an AJC qubit state transition operator $\hat{\overline{A}}$ related to the AJC excitation number operator $\hat{\overline{N}}$ defined by
$$\hat{\overline{A}}=\overline{\delta}s_z+g(\hat{a}s_-+\hat{a}^\dagger s_+) \ ; \quad\quad \hat{\overline{A}}^2=\frac{1}{4}\overline{\delta}^2+g^2(~\hat{\overline{N}}-1) \eqno(7b)$$
we express the effective AJC Hamiltonian $\overline{H}_{AJC}$ in equation (3$e$) in the form
$$\overline{H}_{AJC}=\hbar\omega(~\hat{\overline{N}}-1)+\hbar\hat{\overline{A}} \eqno(7c)$$
Applying $\hat{\overline{A}}$ from equation (7$b$) on the state $\vert\psi_{en}\rangle$ in equation (7$a$), reorganizing, then applying $\hat{\overline{A}}$ on the resulting transition state $\vert~\overline\phi_{en}\rangle$, we determine AJC qubit states $\vert\psi_{en}\rangle$ , $\vert~\overline\phi_{en}\rangle$ satisfying qubit state algebraic operations
$$\hat{\overline{A}}~\vert\psi_{en}\rangle=\overline{R}_{en}\vert~\overline\phi_{en}\rangle \ ; \quad\quad \hat{\overline{A}}~\vert~\overline\phi_{en}\rangle=\overline{R}_{en}\vert\psi_{en}\rangle \eqno(7d)$$
where
$$\vert~\overline{\phi}_{en}\rangle=\overline{c}_{en}\vert\psi_{en}\rangle+\overline{s}_{en}\vert\psi_{gn-1}\rangle \ ; \quad\quad \overline{R}_{en}=g\sqrt{n+(\xi+\varepsilon)^2} \ ; \quad \overline{c}_{en}=\frac{\overline{\delta}}{2\overline{R}_{en}} \ ; \quad \overline{s}_{en}=\frac{g\sqrt{n}}{\overline{R}_{en}} \ ; \quad \varepsilon=\frac{\omega}{g} \eqno(7e)$$
where we have rewritten $\overline{\delta}=\delta+2\omega$ , $\overline{\delta}/2g=\delta/2g + \omega/g$ and the parameter $\xi=\delta/2g$ is defined in equation (5$f$). We identify $\overline{R}_{en}$ as the Rabi frequency of oscillations between the AJC qubit states $\vert\psi_{en}\rangle$ , $\vert~\overline\phi_{en}\rangle$, well developed in [7 , 8] and in subsection~$3.2$ below.

Noting that the qubit states $\vert\psi_{en}\rangle$ , $\vert~\overline\phi_{en}\rangle$ are non-orthogonal satisfying
$$\langle\psi_{en}\vert\psi_{en}\rangle=1 \ , \quad\quad \langle\psi_{en}\vert~\overline\phi_{en}\rangle=\overline{c}_{en} \ , \quad\quad \langle~\overline\phi_{en}\vert\psi_{en}\rangle=\overline{c}_{en} \ , \quad\quad \langle~\overline\phi_{en}\vert~\overline\phi_{en}\rangle=1 \eqno(7f)$$
we introduce normalized AJC eigenstates $\vert~\overline\Psi_{en}^{~+}\rangle$ , $\vert~\overline\Psi_{en}^{~-}\rangle$ obtained as simple linear combinations of the qubit states in the form
$$\vert~\overline\Psi_{en}^{~+}\rangle=\frac{1}{\sqrt{2(1+\overline{c}_{en})}}~(\vert\psi_{en}\rangle+\vert~\overline\phi_{en}\rangle) \ ; \quad\quad \vert~\overline\Psi_{en}^{~-}\rangle=\frac{1}{\sqrt{2(1-\overline{c}_{en})}}~(\vert\psi_{en}\rangle-\vert~\overline\phi_{en}\rangle) \eqno(7g)$$
satisfying eigenvalue equations
$$\hat{\overline{A}}~\vert~\overline\Psi_{en}^{~\pm}\rangle=\pm\overline{R}_{en}~\vert~\overline\Psi_{en}^{~\pm}\rangle \ ; \quad\quad \hat{\overline{N}}~\vert~\overline\Psi_{en}^{~\pm}\rangle=(n+1)\vert~\overline\Psi_{en}^{~\pm}\rangle $$
$$\overline{H}_{AJC}~\vert~\overline\Psi_{en}^{~\pm}\rangle=\overline{E}_{en}^{~\pm}~\vert~\overline\Psi_{en}^{~\pm}\rangle \ ; \quad\quad
\overline{E}_{en}^{~\pm}=\hbar\omega n \pm \hbar\overline{R}_{en} \eqno(7h)$$
If we now set $n=0$ in equations (7$e$) , (7$g$) , (7$h$), the general $n\ge0$ eigenstates $\vert~\overline\Psi_{en}^{~\pm}\rangle$ reduce to the forms (recalling $\overline{\delta}=\omega_0+\omega$~)
$$n=0~:\quad\quad \vert~\overline\Psi_{en}^{~+}\rangle ~\to~ \vert~\overline\Psi_{e0}^{~+}\rangle=\vert\psi_{e0}\rangle \ ; \quad \quad \overline{E}_{e0}^{~+}=\frac{1}{2}\hbar(\omega_0+\omega) $$
$$\vert~\overline\Psi_{en}^{~-}\rangle ~\to~ \vert~\overline\Psi_{e0}^{~-}\rangle=0 \ ; \quad\quad \overline{E}_{e0}^{~-}=-\frac{1}{2}\hbar(\omega_0+\omega) \eqno(7i)$$
which show that $\vert~\overline\Psi_{en}^{~+}\rangle$ is the general $n\ge0$ AJC eigenstate which reduces to the $n=0$ initial state $\vert\psi_{e0}\rangle$ with the correct AJC energy eigenvalue $\overline{E}_{e0}^{~+}=\frac{1}{2}\hbar(\omega_0+\omega)$ agreeing precisely with equation (5$b$). Notice that for $n=0$, the eigenstate $\vert~\overline\Psi_{en}^{~-}\rangle$ reduces to $\vert~\overline\Psi_{e0}^{~-}\rangle=0$ specified by energy eigenvalue $\overline{E}_{e0}^{~-}=-\frac{1}{2}\hbar(\omega_0+\omega)$ which may represent a closed state in the lower AJC spectrum with the atom in the normal ground state of energy $-\frac{1}{2}\hbar\omega_0$ and the field mode in the \emph{antinormal} vacuum state of negative energy $-\frac{1}{2}\hbar\omega$.

From equation (7$i$), we identify $\vert~\overline\Psi_{en}^{~+}\rangle$ in equation (7$g$) as the consistent $n\ge0$ generalization of the AJC eigenstate defining the general initial state for general QRM dynamics generated by the effective JC Hamiltonian $H_{JC}$ in RF which we now present below. In this respect, we substitute the definition of $\vert~\overline\phi_{en}\rangle$ from equation (7$e$) into equation (7$g$) and reorganize to express $\vert~\overline\Psi_{en}^{~+}\rangle$ in the form
$$\vert~\overline\Psi_{en}^{~+}\rangle=
\frac{1}{\sqrt{2(1+\overline{c}_{en})}}~(~(1+\overline{c}_{en})\vert\psi_{en}\rangle+\overline{s}_{en}\vert\psi_{gn-1}\rangle) \eqno(7j)$$
Substituting the definitions of $\vert\psi_{en}\rangle$ , $\vert\psi_{gn-1}\rangle$ from equation (7$a$) reveals that $\vert~\overline\Psi_{en}^{~+}\rangle$ is an entangled state. Note that choosing an AJC eigenstate as the initial state inactivates the AJC interaction in the QRM dynamics in RF, noting that according to the eigenvalue equation (7$h$), $\overline{H}_{AJC}$ generates only plane wave evolution $e^{-\frac{i}{\hbar}E_{en}^+t}\vert~\overline\Psi_{en}^{~+}\rangle$.

The general time evolving state $\vert\Psi_{RF}(t)\rangle$ of general QRM dynamics in RF is generated from the general initial $n\ge0$ AJC eigenstate $\vert~\overline\Psi_{en}^{~+}\rangle$ through the effective JC Hamiltonian $H_{JC}$ according to
$$\vert\Psi_{RF}(t)\rangle=U_{JC}(t)\vert~\overline\Psi_{en}^{~+}\rangle \eqno(8a)$$
where the time evolution operator $U_{JC}(t)$ is defined in equations (6$a$)-(6$b$). Substituting $\vert~\overline\Psi_{en}^{~+}\rangle$ from equation (7$j$) into equation (8$a$) gives the form
$$\vert\Psi_{RF}(t)\rangle=
\frac{1}{\sqrt{2(1+\overline{c}_{en})}}~(~(1+\overline{c}_{en})\vert\Psi_{en}(t)\rangle+\overline{s}_{en}\vert\Psi_{gn-1}(t)\rangle) $$
$$\vert\Psi_{en}(t)\rangle=U_{JC}(t)\vert\psi_{en}\rangle \ ; \quad\quad \vert\Psi_{gn-1}(t)\rangle=U_{JC}(t)\vert\psi_{gn-1}\rangle \eqno(8b)$$
Applying the JC qubit state transition operator $\hat{A}$ from equation (5$c$) on $\vert\psi_{en}\rangle$ , $\vert\psi_{gn-1}\rangle$ generates the respective qubit states $(\vert\psi_{en}\rangle~,~\vert\phi_{en}\rangle)$ , $(\vert\psi_{gn-1}\rangle~,~\vert\phi_{gn-1}\rangle)$ satisfying state transition algebraic operations
$$\hat{A}~\vert\psi_{en}\rangle=R_{en+1}\vert\phi_{en}\rangle \ ; \quad\quad \hat{A}~\vert\phi_{en}\rangle=R_{en+1}\vert\psi_{en}\rangle \ ; \quad\quad \vert\phi_{en}\rangle=c_{en+1}\vert\psi_{en}\rangle+s_{en+1}\vert\psi_{gn+1}\rangle $$
$$\vert\psi_{gn+1}\rangle=\vert gn+1\rangle \ ; \quad\quad R_{en+1}=g\sqrt{(n+1)+\xi^2} \ ; \quad\quad c_{en+1}=\frac{\delta}{2R_{en+1}} \ ; \quad\quad s_{en+1}=\frac{g\sqrt{n+1}}{R_{en+1}} \eqno(8c)$$

$$\hat{A}~\vert\psi_{gn-1}\rangle=R_{gn-1}\vert\phi_{gn-1}\rangle \ ; \quad\ \hat{A}~\vert\phi_{gn-1}\rangle=R_{gn-1}\vert\psi_{gn-1}\rangle \ ; \quad\ \vert\phi_{gn-1}\rangle=-c_{gn-1}\vert\psi_{gn-1}\rangle+s_{gn-1}\vert\psi_{en-2}\rangle $$
$$\vert\psi_{en-2}\rangle=\vert en-2\rangle \ ; \quad\quad R_{gn-1}=g\sqrt{(n-1)+\xi^2} \ ; \quad\quad c_{gn-1}=\frac{\delta}{2R_{gn-1}} \ ; \quad\quad s_{gn-1}=\frac{g\sqrt{n-1}}{R_{gn-1}} \eqno(8d)$$
where $R_{en+1}$ , $R_{gn-1}$ are the respective Rabi frequencies of qubit oscillations.

Substituting $U_{JC}(t)$ from equation (6$b$) into equation (8$b$), noting
$$\hat{N}\vert\psi_{en}\rangle=(n+1)\vert\psi_{en}\rangle \ ; \quad\quad \hat{N}\vert\psi_{gn-1}\rangle=(n-1)\vert\psi_{gn-1}\rangle \eqno(8e)$$
and using equation (6$d$) with repeated application of $\hat{A}$ on $\vert\psi_{en}\rangle$ , $\vert\psi_{gn-1}\rangle$ even and odd number of times, giving general qubit state algebraic relations similar to the relations in equation (5$g$), then introducing trigonometric functions according to the expansions as appropriate, we obtain
$$\vert\Psi_{en}(t)\rangle=e^{-i\omega(n+1)t}(\cos(R_{en+1}t)\vert\psi_{en}\rangle-i\sin(R_{en+1}t)\vert\phi_{en}\rangle) $$
$$\vert\Psi_{gn-1}(t)\rangle=e^{-i\omega(n-1)t}(\cos(R_{gn-1}t)\vert\psi_{gn-1}\rangle-i\sin(R_{gn-1}t)\vert\phi_{gn-1}\rangle) \eqno(8f)$$
Substituting these into equation (8$b$) provides the explicit form of the general time evolving QRM state $\vert\Psi_{RF}(t)\rangle$ in RF generated by the effective JC Hamiltonian $H_{JC}$ from the general $n\ge0$ entangled AJC eigenstate $\vert~\overline\Psi_{en}^{~+}\rangle$ according to equation (8$a$). Introducing the definitions of the states $(~\vert\psi_{en}\rangle ~,~ \vert\psi_{gn+1}\rangle ~,~ \vert\phi_{en}\rangle~)$ and $(~\vert\psi_{gn-1}\rangle ~,~ \vert\psi_{en-2}\rangle ~,~ \vert\phi_{gn-1}\rangle~)$ from equations (7$a$) , (8$c$) , (8$d$) reveals that $\vert\Psi_{RF}(t)\rangle$ is a general time evolving entangled state.

We easily obtain orthonormalization relations
$$\langle\Psi_{en}(t)\vert\Psi_{en}(t)\rangle=1 \ ; \quad \langle\Psi_{en}(t)\vert\Psi_{gn-1}(t)\rangle=0 \ ; \quad \langle\Psi_{gn-1}(t)\vert\Psi_{en}(t)\rangle=0 \ ; \quad \langle\Psi_{gn-1}(t)\vert\Psi_{gn-1}(t)\rangle=1 $$
$$\langle\Psi_{RF}(t)\vert\Psi_{RF}(t)\rangle=1 \eqno(8g)$$
We obtain the JC excitation number $\overline{N}(t)$ in the general time evolving QRM state $\vert\Psi_{RF}(t)\rangle$ in the form
$$\overline{N}(t)=n+1-\frac{\overline{s}_{en}^2}{1+\overline{c}_{en}} \eqno(8h)$$
which once again confirms that the JC excitation number is conserved in RF as expected according to the commutation relation $[~\hat{N}~,~H_{JC}~]=0$ in equation (4$a$). The atomic population inversion and excitation $\overline{s}_z(t)$ , $\overline{s_+s_-}(t)$, the field mode mean photon number $\overline{n}(t)$ and the AJC excitation number $\overline{\overline{N}}(t)$ in the general QRM state $\vert\Psi_{RF}(t)\rangle$ in RF are obtained in the form
$$\overline{s}_z(t)=\frac{1}{4(1+\overline{c}_{en})}\{~(1+\overline{c}_{en})^2(1-2s_{en+1}^2\sin^2(R_{en+1}t))-
\overline{s}_{en}^2(1-2s_{gn-1}^2\sin^2(R_{gn-1}t))~\} $$
$$\overline{s_+s_-}(t)=\frac{1}{2}+\overline{s}_z(t) \ ; \quad\quad \overline{n}(t)=n+\frac{1}{2}-\frac{\overline{s}_{en}^2}{1+\overline{c}_{en}}-\overline{s}_z(t) \ ; \quad\quad \overline{\overline{N}}(t)=n-\frac{\overline{s}_{en}^2}{1+\overline{c}_{en}}+2(1-\overline{s}_z(t)) \eqno(8i)$$
We observe that setting $n=0$ reduces $\vert\Psi_{RF}(t)\rangle$ in equation (8$b$) to $\vert\Psi_{e0}(t)\rangle$ in equation (6$e$) , (6$f$) and the results in equation ((8$i$) reduce to the corresponding results in equation (6$h$), showing that QRM dynamical evolution from the general $n\ge0$ entangled AJC eigenstate $\vert~\overline\Psi_{en}^{~+}\rangle$ in equation (7$j$) is a consistent generalization of the dynamical evolution from the $n=0$ AJC eigenstate $\vert\psi_{e0}\rangle$ in equation (5$a$). Plots of the excitation numbers in equation (8$i$) for initial field mode photon number $n=0$ reproduce the corresponding plots in Fig.$1$-Fig.$4$.

For initial photon numbers $n\ge1$, the time evolution of the excitation numbers in equation (8$i$) is characterized by quantum collapses and revivals largely determined by the field mode initial photon number $n$, which we display for the arbitrarily chosen case $n=40$ in Fig.$5$-Fig.$7$ for the atomic excitation number $X=\overline{s_+s_-}(t)$, field mode mean photon number $F=\overline{n}(t)$ and the AJC excitation number $\overline{\overline{N}}(t)$, respectively. In contrast to the collapse and revival phenomena in the USC-DSC regime due to specification of the atom-field initial state with field mode in initial displaced vacuum state in [9], the collapse and revival phenomenon revealed here in Fig.$5$-Fig.$7$ is due to the superposition of time evolving entangled states $\vert\Psi_{en}(t)\rangle$ , $\vert\Psi_{gn-1}(t)\rangle$ with competing Rabi frequencies $R_{en+1}$ , $R_{gn-1}$ which constitute the general QRM state $\vert\Psi_{RF}(t)\rangle$ in RF according to equations (8$b$) , (8$f$). We observe that the collapses and revivals in Fig.$5$-Fig.$7$ agree particularly well with the field mode mean photon number collapses and revivals obtained in [1].

%\newpage
\begin{figure}[ph]
\centering
\includegraphics[width=0.30\linewidth]{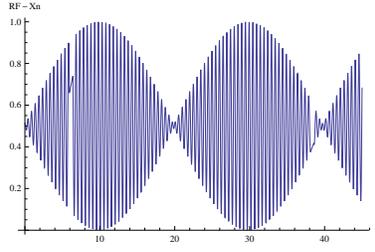}
\caption{JC-atomic excitation number in RF $\overline{s_+s_-}(\tau)~,~\tau=gt :\quad \xi=0~;~\varepsilon=...~;~n=40$}
\label{Fig}
\end{figure}

\begin{figure}[ph]
\centering
\includegraphics[width=0.30\linewidth]{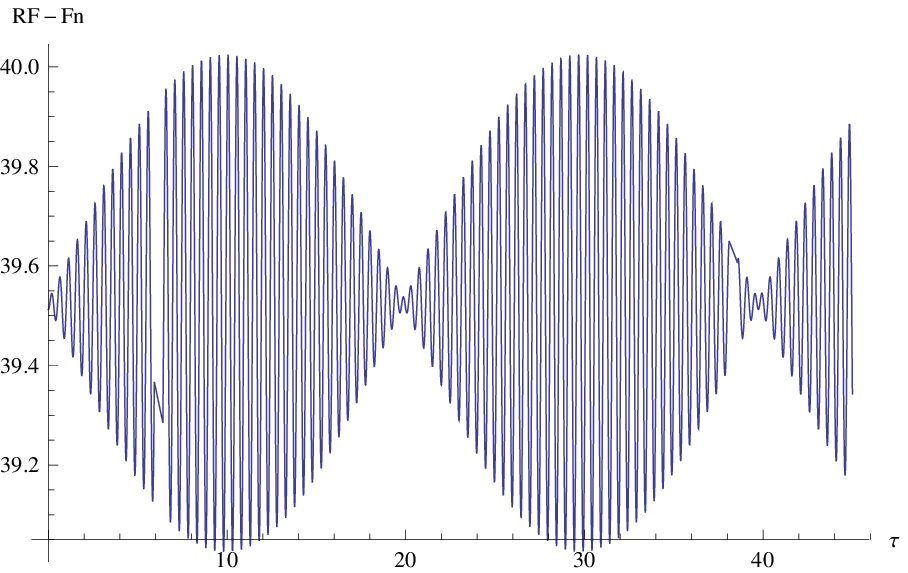}
\caption{JC-field mode mean photon number in RF $\overline{n}(\tau)~,~\tau=gt :\quad \xi=0~;~\varepsilon=...~;~n=40$}
\label{Fig}
\end{figure}

%\begin{figure}[ph]
%\centering
%\includegraphics[width=0.30\linewidth]{QRM-RF-JCEn.eps}
%\caption{JC-excitation number in RF $\overline{N}(\tau)~,~\tau=gt :\quad \xi=0~;~\varepsilon=0.16~;~n=0$}
%\label{Fig}
%\end{figure}

\begin{figure}[ph]
\centering
\includegraphics[width=0.30\linewidth]{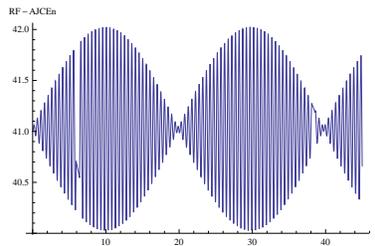}
\caption{AJC-excitation number in RF $\overline{\overline{N}}(\tau)~,~\tau=gt :\quad \xi=0~;~\varepsilon=...~;~n=40$}
\label{Fig}
\end{figure}

\newpage
We note that the collapse-revival phenomenon revealed here in Fig.$5$-Fig.$7$ is a familiar dynamical feature of JC interaction mechanism for atom-field initial states formed from superpositions of atom or field mode states with specified initial photon distributions, which may not be related in any way to the interaction mechanisms in the USC-DSC regimes of QRM as generally defined in [1 , 2 , 4 , 9-15]. QRM dynamics in the USC-DSC regime has generally been associated with the AJC component, but without specifying how the AJC interaction generates dynamical evolution from an initial state, which we now present in the next subsection.

\subsection{QRM dynamics in CRF}
We have established in subsection~$2.3$ that QRM dynamics in CRF is generated by the effective AJC Hamiltonian $\overline{H}_{JC}$ in CRWA according to equations (3$e$)-(3$f$). We note that various theoretical and experimental studies of QRM beyond the RWA have characterized CRF as the USC-DSC regime [1 , 2 , 4 , 9-15]. With this USC-DSC characterization in mind, we follow the experiments [9-12] in identifying and using the state
$\vert g0\rangle$ with the field mode in the vacuum state $\vert 0\rangle$ and the atom in the ground state $\vert g\rangle$ as the appropriate initial state for QRM dynamics in CRF. Here, we establish that the initial state $\vert g0\rangle$ in CRF is an eigenstate of the effective JC Hamiltonian $H_{JC}$ and provide a consistent generalization to the corresponding $n\ge0$ initial JC eigenstate which reduces to $\vert g0\rangle$ for $n=0$.

\subsubsection{Dynamics from initial state $\vert g0\rangle$}
We introduce appropriate notation $\vert\psi_{g0}\rangle$ for the initial state and $\vert\psi_{e1}\rangle$ for the associated transition state defined in standard notation as
$$\vert\psi_{g0}\rangle=\vert g0\rangle \quad\quad ; \quad\quad \vert\psi_{e1}\rangle=\vert e1\rangle \eqno(9a)$$
Determining the dynamical evolution of the QRM ground state $\vert g0\rangle$ has been problematic, noting that the effective JC Hamiltonian $H_{JC}$ cannot generate dynamical evolution of $\vert g0\rangle$ in CRF, since this initial state is an eigenstate of $H_{JC}$. Indeed, using $H_{JC}$ from equation (3$b$) and applying standard atom and field mode state algebraic operations, we easily establish that the QRM ground state $\vert\psi_{g0}\rangle$ is an eigenstate of the effective JC Hamiltonian satisfying an eigenvalue equation (recall $\delta=\omega_0-\omega$)
$$H_{JC}\vert\psi_{g0}\rangle=-\frac{1}{2}\hbar(\omega_0-\omega)\vert\psi_{g0}\rangle \eqno(9b)$$
where we identify the energy eigenvalue $-\frac{1}{2}\hbar(\omega_0-\omega)$ as the atomic ground state energy $-\frac{1}{2}\hbar\omega_0$ and the field mode vacuum state energy $\frac{1}{2}\hbar\omega$ as expected. Equation (9$b$) means that the effective JC Hamiltonian $H_{JC}$ generates only plane wave evolution $e^{\frac{it}{2}(\omega_0-\omega)}\vert\psi_{g0}\rangle$ of the initial state $\vert\psi_{g0}\rangle$ which does not describe the general QRM dynamics in CRF. However, the effective AJC Hamiltonian $\overline{H}_{AJC}$ generates dynamical evolution of the initial state $\vert\psi_{g0}\rangle$ into a time evolving entangled state which describes the general QRM dynamics in CRF as we now demonstrate.

We recall the AJC qubit state transition operator $\hat{\overline{A}}$ and effective Hamiltonian $\overline{H}_{AJC}$ as defined in equations (7$b$)-(7$c$), which we now rewrite here for ease of reference as
$$\hat{\overline{A}}=\overline{\delta}s_z+g(\hat{a}s_-+\hat{a}^\dagger s_+) \ ; \quad\quad \hat{\overline{A}}^2=\frac{1}{4}\overline{\delta}^2+g^2(~\hat{\overline{N}}-1) \ ; \quad\quad \overline{H}_{AJC}=\hbar\omega(~\hat{\overline{N}}-1)+\hbar\hat{\overline{A}} \eqno(9c)$$
Conservation of the AJC excitation number in the dynamics generated by $\overline{H}_{AJC}$ in CRF is easily proved by using the relations in equation (9$c$) to show that $\hat{\overline{N}}$ commutes with $\overline{H}_{AJC}$, thus confirming equations (2$f$) , (4$a$) and here again simplifying the earlier proof in [6].

Applying $\hat{\overline{A}}$ from equation (9$c$) on the initial state $\vert\psi_{g0}\rangle$ defined in equation (9$a$), using standard atom-field state algebraic operations and reorganizing as appropriate, we obtain AJC qubit states $\vert\psi_{g0}\rangle$ , $\vert~\overline\phi_{g0}\rangle$ satisfying qubit state transition algebraic operations in the form
$$\hat{\overline{A}}~\vert\psi_{g0}\rangle=\overline{R}_{g0}\vert~\overline\phi_{g0}\rangle \ ; \quad\quad \hat{\overline{A}}~\vert~\overline\phi_{g0}\rangle=\overline{R}_{g0}\vert\psi_{g0}\rangle \eqno(9d)$$
where $\vert~\overline\phi_{g0}\rangle$ is an entangled qubit transition state obtained in the form
$$\vert~\overline\phi_{g0}\rangle=-\overline{c}_{g0}\vert\psi_{g0}\rangle+\overline{s}_{g0}\vert\psi_{e1}\rangle \ ; \quad\quad \overline{R}_{g0}=g\sqrt{1+(\xi+\varepsilon)^2} \ ; \quad\quad \overline{c}_{g0}=\frac{\overline{\delta}}{2\overline{R}_{g0}} \ ; \quad\quad \overline{s}_{g0}=\frac{g}{\overline{R}_{g0}} \eqno(9e)$$
The general time evolving state $\vert~\overline\Psi_{g0}(t)\rangle$ describing QRM dynamics generated by $\overline{H}_{AJC}$ from the initial state $\vert\psi_{g0}\rangle$ is obtained in the form
$$\vert~\overline\Psi_{g0}(t)\rangle=\overline{U}_{AJC}(t)\vert\psi_{g0}\rangle \ ; \quad\quad \overline{U}_{AJC}(t)=e^{-\frac{it}{\hbar}\overline{H}_{AJC}} \eqno(10a)$$
where $\overline{U}_{AJC}(t)$ is the AJC time evolution operator which on substituting $\overline{H}_{AJC}$ from equation (9$c$) and noting the commutation relation $[~\hat{\overline{N}}~,~\hat{\overline{A}}~]=0$ takes the factorized form
$$\overline{U}_{AJC}(t)=e^{-it\hat{\overline{A}}}e^{-i\omega t(~\hat{\overline{N}}-1)} \eqno(10b)$$
Substituting this into equation (10$a$) and applying $\hat{\overline{N}}=\hat{a}\hat{a}^\dagger+s_-s_+$ on $\vert\psi_{g0}\rangle$ gives
$$(~\hat{\overline{N}}-1)\vert\psi_{g0}\rangle=\vert\psi_{g0}\rangle \quad\quad\Rightarrow\quad\quad
\vert~\overline\Psi_{g0}(t)\rangle=e^{-i\omega t}e^{-it\hat{\overline{A}}}~\vert\psi_{g0}\rangle \eqno(10c)$$
Expanding $e^{-it\hat{\overline{A}}}$ in even and odd power terms similar to the corresponding JC time evolution operator expansion in equation (6$d$) and substituting into equation (10$c$), we apply $\hat{\overline{A}}$ on $\vert\psi_{g0}\rangle$ even and odd number of times using the qubit state transition algebraic operations from equation (9$d$) giving relations similar to equation (5$g$) and then introduce trigonometric functions in the expansions as appropriate to obtain the general time evolving state in the final form
$$\vert~\overline\Psi_{g0}(t)\rangle=e^{-i\omega t}(\cos(~\overline{R}_{g0}t)\vert\psi_{g0}\rangle-i\sin(~\overline{R}_{g0}t)\vert~\overline\phi_{g0}\rangle) \eqno(10d)$$
which describes Rabi oscillations at frequency $\overline{R}_{g0}$ between the initial separable state $\vert\psi_{g0}\rangle$ and the entangled transition state $\vert~\overline\phi_{g0}\rangle$. Substituting $\vert~\overline\phi_{g0}\rangle$ from equation (9$e$) into equation (10$d$), reorganizing and introducing the definitions of $\vert\psi_{g0}\rangle$ , $\vert\psi_{e1}\rangle$ from equation (9$a$) reveals that in general, the time evolving state $\vert~\overline\Psi_{g0}(t)\rangle$ is a normalized entangled state obtained in the form
$$\vert~\overline\Psi_{g0}(t)\rangle=e^{-i\omega t}(~(\cos(~\overline{R}_{g0}t)+i~\overline{c}_{g0}\sin(~\overline{R}_{g0}t)~)\vert g0\rangle -i~\overline{s}_{g0}\sin(~\overline{R}_{g0}t)\vert e1\rangle) \ ; \quad\quad \langle~\overline\Psi_{g0}(t)\vert~\overline\Psi_{g0}(t)\rangle=1 \eqno(10e)$$
Hence, as we set out to demonstrate, the effective AJC Hamiltonian $\overline{H}_{AJC}$ generates dynamical evolution of the initial atom-field state
$\vert g0\rangle$ into a time evolving entangled state $\vert~\overline\Psi_{g0}(t)\rangle$ in CRF. We observe that this form of dynamical evolution of the QRM ground state $\vert g0\rangle$ into a time evolving entangled state generated by the effective AJC Hamiltonian has never been determined in the various theoretical models or related experiments in [1 , 2 , 4 , 9-15] and others, noting that the solution procedure for AJC interaction has only been developed by the present author in recent work [6-8]. As we pointed out earlier, considering the dynamical features of QRM in the USC-DSC regime, the authors of studies in [1 , 2 , 4 , 9-15] defined the initial atom-field ground state $\vert g0\rangle$ in a more general form with the field mode in initial displaced vacuum state and the atom in initial eigenstate of the spin operator $\sigma_x=s_-+s_+$, so that the ground state evolves into a Schroedinger cat state in the USC-DSC regime [9 , 12 , 14 , 15]. Comparison of QRM dynamics described by $\vert~\overline\Psi_{g0}(t)\rangle$ in CRF with the dynamical features observed in the USC-DSC regime in the QRM simulation experiments in [9-12] may therefore be inappropriate, since the definition of CRF as a dynamical frame of QRM as we have developed it in this article is independent of the coupling parameter $\frac{g}{\omega}$ used to characterize the QRM coupling regimes in [1 , 2 , 4 , 9-15], meaning that CRF may not be equivalent to the USC-DSC regime. With this in mind, we determine the atomic excitation number, field mode mean photon number, the JC and AJC excitation numbers to study QRM dynamical features in the state $\vert\Psi_{g0}(t)\rangle$ in CRF.

Applying the AJC excitation number operator $\hat{\overline{N}}$ on $\vert~\overline\Psi_{g0}(t)\rangle$ gives an eigenvalue equation
$$\hat{\overline{N}}\vert~\overline\Psi_{g0}(t)\rangle=2\vert~\overline\Psi_{g0}(t)\rangle \quad\quad\Rightarrow\quad\quad \overline{\overline{N}}(t)=2 \eqno(10f)$$
from which it follows that the AJC excitation number $\overline{\overline{N}}(t)$ is conserved in the QRM dynamics generated by the effective AJC Hamiltonian $\overline{H}_{AJC}$ in CRF as expected from the corresponding commutation relation $[~\hat{\overline{N}}~,~\overline{H}_{JC}~]=0$ in equation (4$a$). This property has never been investigated experimentally, since, as we have explained earlier, the AJC excitation number operator has largely been unknown to both theoreticians and experimentalists, a fact which has necessitated the work presented in this article.

Noting that AJC operators are defined in antinormal order form, we determine the atomic population inversion $\overline{s}_z(t)$ and antinormal order excitation number $\overline{s_-s_+}(t)$, the field mode antinormal order photon number $\overline{aa^*}(t)=\overline{n}(t)+1$ and the JC excitation number $\overline{N}(t)$ (normal order) in the QRM time evolving state $\vert~\overline\Psi_{g0}(t)\rangle$ in CRF in the form (recall $s_+s_-=\frac{1}{2}+s_z ~;~ s_-s_+=\frac{1}{2}-s_z~;~\hat{a}\hat{a}^\dagger=\hat{a}^\dagger\hat{a}+1$)
$$\overline{s}_z(t)=-\frac{1}{2}(1-2\overline{s}_{g0}^2\sin^2(~\overline{R}_{g0}t)) \ ; \quad\ \overline{s_-s_+}(t)=\frac{1}{2}-\overline{s}_z(t) \ ; \quad\ \overline{n}(t)=\overline{s}_{g0}^2\sin^2(~\overline{R}_{g0}t) \ ; \quad\ \overline{aa^*}(t)=1+\overline{n}(t)$$
$$\overline{N}(t)=\overline{n}(t)+\frac{1}{2}+\overline{s}_z(t) \eqno(10g)$$
It follows from equation (10$g$) that the JC excitation number $\overline{N}(t)$ is non-conserved and evolves in time in the QRM dynamics generated by the effective AJC Hamiltonian $\overline{H}_{AJC}$ in CRF as expected from the corresponding commutation relation $[~\hat{N}~,~\overline{H}_{AJC}~]\ne0$ in equation (4$a$). Experiments [9] have established that the JC excitation number $\overline{N}(t)$ is conserved in the JC regime corresponding to RF here, but evolves in time in the USC-DSC regime, which is associated with, but not necessarily equivalent to CRF. In general, the experimental observations agree with the results we have obtained in equations (6$g$) , (8$h$) in RF and here in equation (10$g$) in CRF. Here, we now only mention the familiar and overemphasized fact that the non-conservation of the AJC excitation number $\overline{\overline{N}}(t)$ in equations (6$h$) , (8$i$) in RF and its conservation determined here in equation (10$f$) in CRF has not been investigated in experiments.

We have plotted the antinormal atomic excitation number $X=\overline{s_-s_+}(t)$, the field mode mean antinormal photon number $F=\overline{aa^*}(t)$, the AJC and JC excitation numbers $\overline{\overline{N}}(t)$ , $\overline{N}(t)$ from equation (10$g$) in Fig.$8$-Fig.$11$, respectively. We notice the striking similarity with the time evolution of corresponding quantities determined in the JC interaction in RF plotted in Fig.$1$-Fig.$4$. The similarity in the form of time evolution is that the excitation and mean photon numbers are defined in quadratic form in both JC and AJC interactions. We observe that the non-quadratic atomic state population inversion $\overline{s}_z(t)$ and the coherence functions $\overline{s}_x$ , $\overline{s}_y(t)$ obtained in JC and AJC interactions in RF , CRF evolve in time in reverse order, which we have not plotted.

%\newpage
\begin{figure}[ph]
\centering
\includegraphics[width=0.30\linewidth]{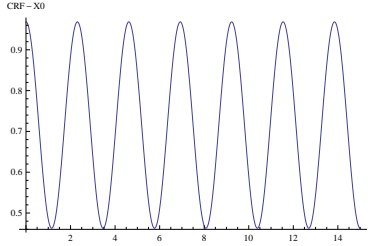}
\caption{AJC-atomic antinormal excitation number in CRF $\overline{s_-s_+}(\tau)~,~\tau=gt :\quad \xi=\frac{1}{1.31}~;~\varepsilon=0.16~;~n=0$}
\label{Fig}
\end{figure}

\begin{figure}[ph]
\centering
\includegraphics[width=0.30\linewidth]{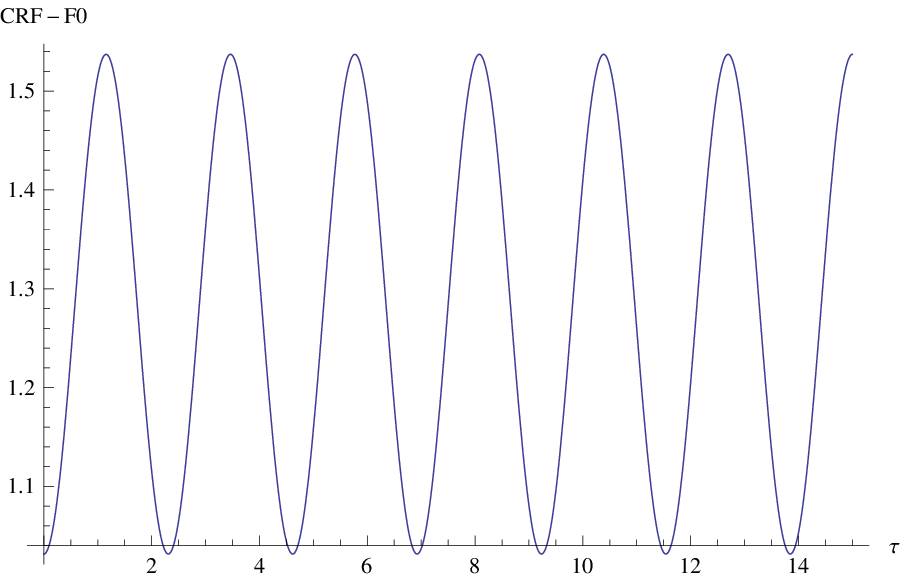}
\caption{AJC-field mode mean antinormal photon number in CRF $\overline{aa^*}(\tau)~,~\tau=gt :\quad \xi=\frac{1}{1.31}~;~\varepsilon=0.16~;~n=0$}
\label{Fig}
\end{figure}

\newpage

\begin{figure}[ph]
\centering
\includegraphics[width=0.30\linewidth]{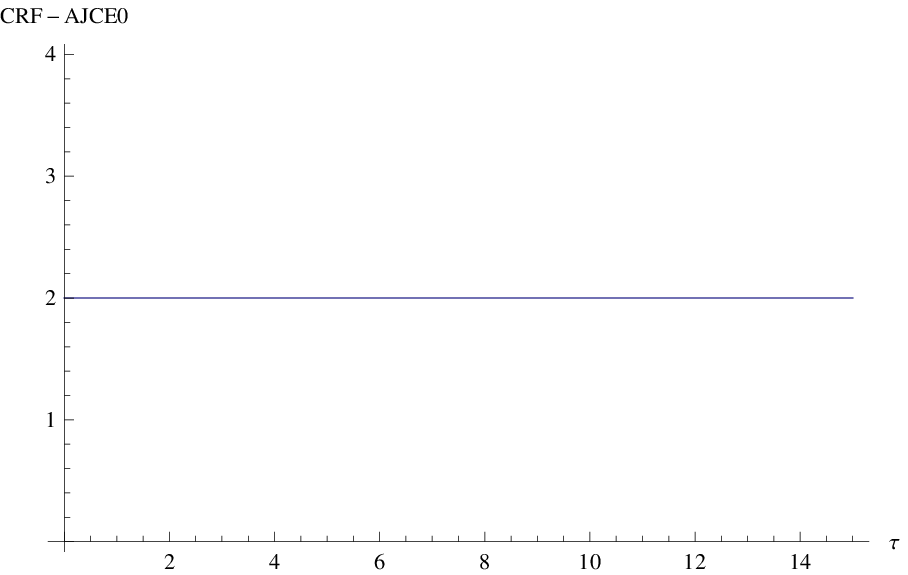}
\caption{AJC-excitation number in CRF $\overline{\overline{N}}(\tau)~,~\tau=gt :\quad \xi=\frac{1}{1.31}~;~\varepsilon=0.16~;~n=0$}
\label{Fig}
\end{figure}

\begin{figure}[ph]
\centering
\includegraphics[width=0.30\linewidth]{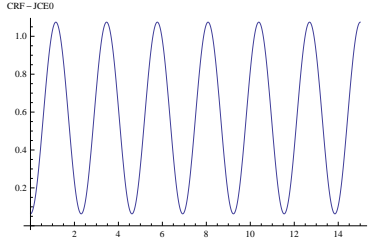}
\caption{JC-excitation number in CRF $\overline{N}(\tau)~,~\tau=gt :\quad \xi=\frac{1}{1.31}~;~\varepsilon=0.16~;~n=0$}
\label{Fig}
\end{figure}
We now generalize the initial atom-field state by determining the $n\ge0$ JC eigenstate which reduces to the ground state $\vert g0\rangle$ at $n=0$.

\subsubsection{Dynamics from a general initial JC eigenstate}
Noting that the initial state $\vert\psi_{00}\rangle$ used above in developing QRM dynamics in CRF is an eigenstate of the JC Hamiltonian $H_{JC}$ according to equation (9$b$), we now provide a consistent generalization to an $n\ge0$ initial JC eigenstate. Since the atom starts in the ground state $\vert g\rangle$, the basic $n\ge0$ atom-field state is $\vert gn\rangle$. Considering that the state algebraic operation for determining a general eigenstate of the JC Hamiltonian couples the state $\vert gn\rangle$ to the state $\vert en-1\rangle$, we introduce appropriate notation $\vert\psi_{gn}\rangle$ , $\vert\psi_{en-1}\rangle$ for the two states in the form
$$\vert\psi_{gn}\rangle=\vert gn\rangle \ ; \quad\quad \vert\psi_{en-1}\rangle=\vert en-1\rangle \eqno(11a)$$
Applying the JC qubit state transition operator $\hat{A}$ from equation (5$c$) on the state $\vert\psi_{gn}\rangle$ in equation (11$a$), reorganizing, then applying $\hat{A}$ on the resulting transition state $\vert\phi_{gn}\rangle$, we determine JC qubit states $\vert\psi_{gn}\rangle$ , $\vert\phi_{gn}\rangle$ satisfying qubit state algebraic operations
$$\hat{A}~\vert\psi_{gn}\rangle=R_{gn}\vert\phi_{gn}\rangle \ ; \quad\quad \hat{A}~\vert\phi_{gn}\rangle=R_{gn}\vert\psi_{gn}\rangle \eqno(11b)$$
where
$$\vert\phi_{gn}\rangle=-c_{gn}\vert\psi_{gn}\rangle+s_{gn}\vert\psi_{en-1}\rangle \ ; \quad\quad R_{gn}=g\sqrt{n+\xi^2} \ ; \quad c_{gn}=\frac{\delta}{2R_{gn}} \ ; \quad s_{gn}=\frac{g\sqrt{n}}{R_{gn}} \eqno(11c)$$
Noting that the qubit states $\vert\psi_{gn}\rangle$ , $\vert\phi_{gn}\rangle$ are non-orthogonal satisfying
$$\langle\psi_{gn}\vert\psi_{gn}\rangle=1 \ , \quad\quad \langle\psi_{gn}\vert\phi_{gn}\rangle=-c_{gn} \ , \quad\quad \langle\phi_{gn}\vert\psi_{gn}\rangle=-c_{gn} \ , \quad\quad \langle\phi_{gn}\vert\phi_{gn}\rangle=1 \eqno(11d)$$
we introduce normalized JC eigenstates $\vert\Psi_{gn}^{~+}\rangle$ , $\vert\Psi_{gn}^{~-}\rangle$ obtained as simple linear combinations of the qubit states in the form
$$\vert\Psi_{gn}^{~+}\rangle=\frac{1}{\sqrt{2(1-c_{gn})}}~(\vert\psi_{gn}\rangle+\vert\phi_{gn}\rangle) \ ; \quad\quad \vert\Psi_{gn}^{~-}\rangle=\frac{1}{\sqrt{2(1+c_{gn})}}~(\vert\psi_{gn}\rangle-\vert\phi_{gn}\rangle) \eqno(11e)$$
satisfying eigenvalue equations
$$\hat{A}~\vert\Psi_{gn}^{~\pm}\rangle=\pm R_{gn}~\vert\Psi_{gn}^{~\pm}\rangle \ ; \quad\quad \hat{N}~\vert\Psi_{gn}^{~\pm}\rangle=n\vert\Psi_{gn}^{~\pm}\rangle $$
$$H_{JC}~\vert\Psi_{gn}^{~\pm}\rangle=E_{gn}^{~\pm}~\vert\Psi_{gn}^{~\pm}\rangle \ ; \quad\quad E_{gn}^{~\pm}=\hbar\omega n \pm \hbar R_{gn} \eqno(11f)$$
If we now set $n=0$ in equations (11$c$) , (11$e$) , (11$f$), the general $n\ge0$ eigenstates $\vert\Psi_{gn}^{~\pm}\rangle$ reduce to the forms (recalling $\delta=\omega_0-\omega$~)
$$n=0~:\quad\quad \vert\Psi_{gn}^{~+}\rangle ~\to~ \vert\Psi_{g0}^{~+}\rangle=0 \ ; \quad \quad E_{g0}^{~+}=\frac{1}{2}\hbar(\omega_0-\omega) $$
$$\vert\Psi_{gn}^{~-}\rangle ~\to~ \vert\Psi_{g0}^{~-}\rangle=\vert\psi_{g0}\rangle \ ; \quad\quad E_{g0}^{~-}=-\frac{1}{2}\hbar(\omega_0-\omega) \eqno(11g)$$
which show that $\vert\Psi_{gn}^{~-}\rangle$ is the general $n\ge0$ JC eigenstate which reduces to the $n=0$ initial state $\vert\psi_{g0}\rangle$ with the correct $JC$ energy eigenvalue $E_{g0}^{~-}=-\frac{1}{2}\hbar(\omega_0-\omega)$ agreeing precisely with equation (9$b$). Notice that for $n=0$, the eigenstate $\vert\Psi_{gn}^{~+}\rangle$ reduces to $\vert\Psi_{g0}^{~+}\rangle=0$ specified by energy eigenvalue $E_{g0}^{~+}=\frac{1}{2}\hbar(\omega_0-\omega)$ which may represent a closed state in the upper JC spectrum with the atom in the normal excited state of energy $\frac{1}{2}\hbar\omega_0$ and the field mode in the \emph{antinormal} vacuum state of negative energy $-\frac{1}{2}\hbar\omega$.

From equation (11$g$), we identify $\vert\Psi_{gn}^{~-}\rangle$ in equation (11$e$) as the consistent $n\ge0$ generalization of the JC eigenstate defining the general initial state for general QRM dynamics generated by the effective AJC Hamiltonian $\overline{H}_{AJC}$ in CRF which we now present below. In this respect, we substitute the definition of $\vert\phi_{gn}\rangle$ from equation (11$c$) into equation (11$e$) and reorganize to express $\vert\Psi_{gn}^{~-}\rangle$ in the form
$$\vert\Psi_{gn}^{~-}\rangle=\frac{1}{\sqrt{2(1+c_{gn})}}~(~(1+c_{gn})\vert\psi_{gn}\rangle-s_{gn}\vert\psi_{en-1}\rangle) \eqno(11h)$$
Substituting the definitions of $\vert\psi_{gn}\rangle$ , $\vert\psi_{en-1}\rangle$ from equation (11$a$) reveals that $\vert\Psi_{gn}^{~-}\rangle$ is an entangled state. Note that choosing a JC eigenstate as the initial state inactivates the JC interaction in the QRM dynamics in CRF, seeing that according to the eigenvalue equation (11$f$), $H_{JC}$ only generates plane wave evolution $e^{-\frac{i}{\hbar}E_{gn}^-t}\vert\Psi_{gn}^-\rangle$.

The general time evolving state $\vert~\overline\Psi_{CRF}(t)\rangle$ of general QRM dynamics in CRF is generated from the general initial $n\ge0$ JC eigenstate $\vert\Psi_{gn}^{~-}\rangle$ through the effective AJC Hamiltonian $\overline{H}_{AJC}$ according to
$$\vert~\overline\Psi_{CRF}(t)\rangle=\overline{U}_{AJC}(t)\vert\Psi_{gn}^{~-}\rangle \eqno(12a)$$
where the time evolution operator $\overline{U}_{AJC}(t)$ is defined in equations (10$a$)-(10$b$). Substituting $\vert\Psi_{gn}^{~-}\rangle$ from equation (11$h$) into equation (12$a$) gives the form
$$\vert~\overline\Psi_{CRF}(t)\rangle=
\frac{1}{\sqrt{2(1+c_{gn})}}~(~(1+c_{gn})\vert~\overline\Psi_{gn}(t)\rangle-s_{gn}\vert~\overline\Psi_{en-1}(t)\rangle) $$
$$\vert~\overline\Psi_{gn}(t)\rangle=\overline{U}_{AJC}(t)\vert\psi_{gn}\rangle \ ; \quad\quad \vert~\overline\Psi_{en-1}(t)\rangle=\overline{U}_{AJC}(t)\vert\psi_{en-1}\rangle \eqno(12b)$$
Applying the AJC qubit state transition operator $\hat{\overline{A}}$ from equation (9$c$) on $\vert\psi_{gn}\rangle$ , $\vert\psi_{en-1}\rangle$ generates the respective qubit states $(\vert\psi_{gn}\rangle~,~\vert~\overline\phi_{gn}\rangle)$ , $(\vert\psi_{en-1}\rangle~,~\vert~\overline\phi_{en-1}\rangle)$ satisfying state transition algebraic operations
$$\hat{\overline{A}}~\vert\psi_{gn}\rangle=\overline{R}_{en+1}\vert~\overline\phi_{gn}\rangle \ ; \quad\quad \hat{\overline{A}}~\vert~\overline\phi_{gn}\rangle=\overline{R}_{gn+1}\vert\psi_{gn}\rangle \ ; \quad\quad \vert~\overline\phi_{gn}\rangle=-\overline{c}_{gn+1}\vert\psi_{gn}\rangle+\overline{s}_{gn+1}\vert\psi_{en+1}\rangle $$
$$\vert\psi_{en+1}\rangle=\vert en+1 \ ; \quad\quad \overline{R}_{gn+1}=g\sqrt{(n+1)+(\xi+\varepsilon)^2} \ ; \quad\quad \overline{c}_{gn+1}=\frac{\overline{\delta}}{2\overline{R}_{gn+1}} \ ; \quad\quad \overline{s}_{gn+1}=\frac{g\sqrt{n+1}}{\overline{R}_{gn+1}} \eqno(12c)$$

$$\hat{\overline{A}}~\vert\psi_{en-1}\rangle=\overline{R}_{en-1}\vert~\overline\phi_{en-1}\rangle \ ; \quad\ \hat{\overline{A}}~\vert~\overline\phi_{en-1}\rangle=\overline{R}_{en-1}\vert\psi_{en-1}\rangle \ ; \quad\ \vert~\overline\phi_{en-1}\rangle=\overline{c}_{en-1}\vert\psi_{en-1}\rangle+\overline{s}_{en-1}\vert\psi_{gn-2}\rangle $$
$$\vert\psi_{gn-2}\rangle=\vert gn-2\rangle \ ; \quad\quad \overline{R}_{en-1}=g\sqrt{(n-1)+(\xi+\varepsilon)^2} \ ; \quad\quad \overline{c}_{en-1}=\frac{\overline{\delta}}{2\overline{R}_{en-1}} \ ; \quad\quad \overline{s}_{en-1}=\frac{g\sqrt{n-1}}{\overline{R}_{en-1}} \eqno(12d)$$
where $\overline{R}_{gn+1}$ , $\overline{R}_{en-1}$ are the respective Rabi frequencies of qubit oscillations.

Substituting $\overline{U}_{AJC}(t)$ from equation (10$b$) into equation (12$b$), noting
$$(~\hat{\overline{N}}-1)~\vert\psi_{gn}\rangle=(n+1)\vert\psi_{gn}\rangle \ ; \quad\quad (~\hat{\overline{N}}-1)\vert\psi_{en-1}\rangle=(n-1)\vert\psi_{en-1}\rangle \eqno(12e)$$
and expanding $e^{-it\hat{\overline{A}}}$ in even and odd power terms similar to the corresponding JC time evolution operator expansion in equation (6$d$), then substituting into equation (12$b$), we apply $\hat{\overline{A}}$ on $\vert\psi_{gn}\rangle$ , $\vert\psi_{en-1}\rangle$ even and odd number times using the qubit state transition algebraic operations from equations (12$c$) , (12$d$) giving relations similar to equation (5$g$) and introduce trigonometric functions in the expansions as appropriate to obtain
$$\vert~\overline\Psi_{gn}(t)\rangle=e^{-i\omega(n+1)t}(\cos(~\overline{R}_{gn+1}t)\vert\psi_{gn}\rangle -
i\sin(~\overline{R}_{gn+1}t)\vert~\overline\phi_{gn}\rangle) $$
$$\vert~\overline\Psi_{en-1}(t)\rangle=e^{-i\omega(n-1)t}(\cos(~\overline{R}_{en-1}t)\vert\psi_{en-1}\rangle -
i\sin(~\overline{R}_{en-1}t)\vert~\overline\phi_{en-1}\rangle) \eqno(12f)$$
Substituting these into equation (12$b$) provides the explicit form of the general time evolving QRM state $\vert~\overline\Psi_{CRF}(t)\rangle$ in CRF generated by the effective AJC Hamiltonian $\overline{H}_{AJC}$ from the general $n\ge0$ initial entangled JC eigenstate $\vert\Psi_{gn}^{~-}\rangle$. Introducing the definitions of the states $(~\vert\psi_{gn}\rangle ~,~ \vert\psi_{en+1}\rangle ~,~ \vert~\overline\phi_{gn}\rangle~)$ and $(~\vert\psi_{en-1}\rangle ~,~ \vert\psi_{gn-2}\rangle ~,~ \vert~\overline\phi_{en-1}\rangle~)$ from equations (11$a$) , (12$c$) , (12$d$) reveals that the general time evolving QRM state $\vert~\overline\Psi_{CRF}(t)\rangle$ in CRF is a time evolving entangled state.

We easily obtain orthonormalization relations
$$\langle~\overline\Psi_{gn}(t)\vert~\overline\Psi_{gn}(t)\rangle=1 \ ; \quad \langle~\overline\Psi_{gn}(t)\vert~\overline\Psi_{en-1}(t)\rangle=0 \ ; \quad \langle~\overline\Psi_{en-1}(t)\vert~\overline\Psi_{gn}(t)\rangle=0 \ ; \quad \langle~\overline\Psi_{en-1}(t)\vert~\overline\Psi_{en-1}(t)\rangle=1 $$
$$\langle~\overline\Psi_{CRF}(t)\vert~\overline\Psi_{CRF}(t)\rangle=1 \eqno(12g)$$
We obtain the AJC excitation number $\overline{\overline{N}}(t)$ in the general time evolving QRM state $\vert~\overline\Psi_{CRF}(t)\rangle$ in the form
$$\overline{\overline{N}}(t)=n+2-\frac{s_{gn}^2}{1+c_{gn}} \eqno(12h)$$
which once again confirms that the AJC excitation number is conserved in CRF as expected according to the commutation relation $[~\hat{\overline{N}}~,~\overline{H}_{AJC}~]=0$ in equation (4$a$). The atomic population inversion and antinormal order excitation $\overline{s}_z(t)$ , $\overline{s_-s_+}(t)$, the field mode mean antinormal order photon number $\overline{aa^*}(t)$ and the JC excitation number $\overline{N}(t)$ (normal order) in the QRM state $\vert~\overline\Psi_{CRF}(t)\rangle$ are obtained in the form
$$\overline{s}_z(t)=-\frac{1}{4(1+c_{gn})}\{~(1+c_{gn})^2(1-2\overline{s}_{gn+1}^2\sin^2(~\overline{R}_{gn+1}t))-
s_{gn}^2(1-2\overline{s}_{en-1}^2\sin^2(~\overline{R}_{en-1}t))~\} $$
$$\overline{s_-s_+}(t)=\frac{1}{2}-\overline{s}_z(t) \ ; \quad\quad \overline{aa^*}(t)=n+\frac{3}{2}-\frac{s_{gn}^2}{1+c_{gn}}+\overline{s}_z(t) \ ; \quad\quad \overline{N}(t)=n+1-\frac{s_{gn}^2}{1+c_{gn}}+2~\overline{s}_z(t) \eqno(12i)$$
We observe that setting $n=0$ reduces $\vert~\overline\Psi_{CRF}(t)\rangle$ in equation (12$b$) to $\vert\Psi_{g0}(t)\rangle$ in equation (10$d$) , (10$e$) and the results in equation ((12$i$) reduce to the corresponding results in equation (10$g$), showing that QRM dynamical evolution from the general $n\ge0$ entangled JC eigenstate $\vert\Psi_{gn}^{~-}\rangle$ in equation (11$h$) is a consistent generalization of the dynamical evolution from the $n=0$ JC eigenstate $\vert\psi_{g0}\rangle$ in equation (9$a$). Plots of the mean values in equation (12$i$) for $n=0$ reproduce the corresponding plots in Fig.$8$-Fig.$11$, while plots for field mode initial photon numbers $n\ge1$ show time evolution characterized by quantum collapses and revivals largely determined by the initial field mode photon number $n$.

We have plotted the atomic antinormal excitation number $X=\overline{s_-s_+}(t)$ , field mode mean antinormal photon number $F=\overline{aa^*}(t)$ and the JC excitation number $\overline{N}(t)$ for arbitrarily chosen initial photon number $n=40$ and dimensionless parameters $\xi$ , $\varepsilon$, in Fig.$12$ , Fig.$13$ , Fig.$14$, respectively, which clearly undergo collapses and revivals similar to the corresponding JC cases in RF presented in Fig.$5$-Fig.$7$. Again, we observe that the field mode mean photon number collapse-revival profile takes the form obtained in the full QRM DSC regime in [1]. Here again, we consider the collapse-revival phenomenon to be a dynamical feature of AJC interaction mechanism for atom-field initial superposition state such as the $n\ge0$ entangled JC eigenstate, noting that CRF where the effective AJC interaction is dominant is not necessarily equivalent to the USC-DSC regime of QRM as usually defined.

%\newpage
\begin{figure}[ph]
\centering
\includegraphics[width=0.30\linewidth]{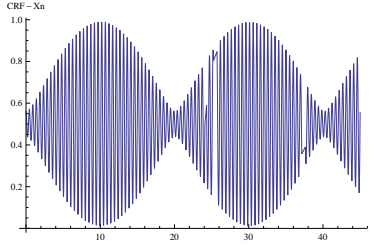}
\caption{AJC-atomic antinormal excitation number in CRF $\overline{s_-s_+}(\tau)~,~\tau=gt :\quad \xi=\frac{1}{1.31}~;~\varepsilon=0.16~;~n=40$}
\label{Fig}
\end{figure}

\begin{figure}[ph]
\centering
\includegraphics[width=0.30\linewidth]{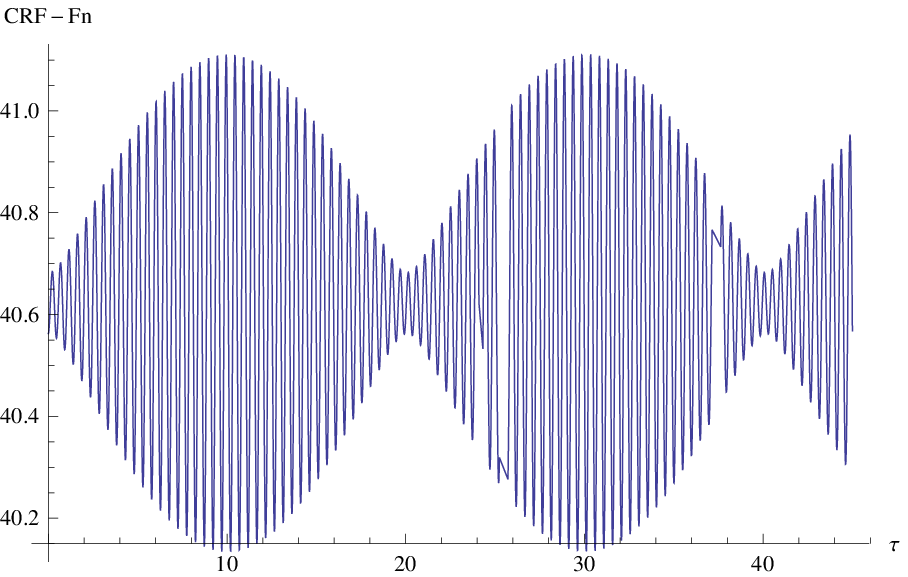}
\caption{AJC-field mode mean antinormal photon number in CRF $\overline{aa^*}(\tau)~,~\tau=gt :\quad \xi=\frac{1}{1.31}~;~\varepsilon=0.16~;~n=40$}
\label{Fig}
\end{figure}

%\begin{figure}[ph]
%\centering
%\includegraphics[width=0.30\linewidth]{QRM-CRF-AJCEn.eps}
%\caption{AJC-excitation number in CRF $\overline{\overline{N}}(\tau)~,~\tau=gt :\quad \xi=\frac{1}{1.31}~;~\varepsilon=0.16~;~n=0$}
%\label{Fig}
%\end{figure}

\begin{figure}[ph]
\centering
\includegraphics[width=0.30\linewidth]{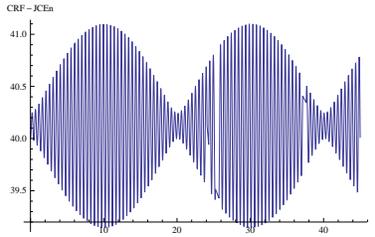}
\caption{JC-excitation number in CRF $\overline{N}(\tau)~,~\tau=gt :\quad \xi=\frac{1}{1.31}~;~\varepsilon=0.16~;~n=40$}
\label{Fig}
\end{figure}

\newpage

\section{Conclusion}
By demonstrating that the AJC interaction has a conserved excitation number operator and is exactly solvable, we have addressed a major challenge of theoretical and experimental efforts to investigate the internal dynamics of QRM. We have established that QRM has two correlated dynamical frames; the rotating frame (RF) where the dynamics is dominated by the exactly solved JC interaction characterized by red-sideband transitions, with a conserved JC excitation number operator which generates the $U(1)$ symmetry of RF, and, the counter-rotating frame (CRF) where the dynamics is dominated by the exactly solved AJC interaction characterized by blue-sideband transitions, with a conserved AJC excitation number operator which generates the $U(1)$ symmetry of CRF. The two conserved, JC and AJC, excitation number operators commute and generate a common parity symmetry operator of both JC and AJC interactions, thereby providing the parity symmetry operator of the full QRM. The $U(1)$ symmetry operator of JC reduces QRM Hamiltonian to an effective JC Hamiltonian in an RWA in RF, while the $U(1)$ symmetry operator of AJC reduces QRM Hamiltonian to an effective AJC Hamiltonian in a CRWA in CRF. Considering the initial atom-field states $\vert e0\rangle$ and $\vert g0\rangle$, preferred as the fundamental QRM initial states in the experiments, we have established that the effective JC Hamiltonian $H_{JC}$ generates dynamical evolution of the state $\vert e0\rangle$ into a time evolving entangled state in RF, while the effective AJC Hamiltonian $\overline{H}_{AJC}$ generates dynamical evolution of the (absolute) ground state $\vert g0\rangle$ into a time evolving entangled state in CRF, thus addressing another major challenge of determining QRM dynamics beyond RWA. Identifying the initial atom-field states $\vert e0\rangle$ and $\vert g0\rangle$ as eigenstates of the effective AJC and JC Hamiltonians, respectively, we have derived the corresponding general $n\ge0$ entangled AJC and JC eigenstates as consistent generalizations of QRM initial states in RF under RWA and CRF under CRWA. The general QRM state in RF or CRF is then a general time evolving entangled state generated by $H_{JC}$ from the general $n\ge0$ initial entangled AJC eigenstate $\vert~\overline\Psi_{en}^{~+}\rangle$ or a general time evolving entangled state generated by $\overline{H}_{AJC}$ from the general $n\ge0$ initial entangled JC eigenstate $\vert\Psi_{gn}^{~-}\rangle$. In QRM dynamics from the general $n\ge0$ initial entangled states in RF or CRF, the general time evolution of the atomic population inversion and excitation number, the field mode mean photon number and the JC/AJC excitation numbers undergo quantum collapses and revivals determined by the initial field mode photon numbers $n\ge1$, where we note that the JC excitation number is conserved in RF, but evolves in time in CRF, while the AJC excitation number is conserved in CRF, but evolves in time in RF. An important point which arises is that the clear specification of the QRM dynamical frames RF and CRF dominated by the exactly solved effective JC and AJC interaction mechanisms, respectively, now calls to question the true physical interpretation of the coupling regimes, which have been characterized in the theoretical models and experimental designs as the weak-strong coupling (WSC) regime where JC interaction dominates and the USC-DSC regime where AJC interaction is believed to be dominant. Considering the QRM dynamical frames as we have specified and demonstrated their physical characteristics in the present article, can we consistently interpret the WSC regime as RF dominated by the JC interaction mechanism and the USC-DSC regime as CRF dominated by the AJC interaction mechanism ? Such an interpretation may have to be reviewed, noting that the basic definitions of RF/CRF do not depend explicitly on the dimensionless coupling parameter $\frac{g}{\omega}$ used to characterize the coupling regimes WSC , USC-DSC and that the respective JC and AJC qubit state transition algebraic operations in equations (5$e$-5$f$~,~8$c$-8$d$) which characterize dynamics in RF and (9$d$-9$e$~,~12$c$-12$d$) which characterize dynamics in CRF are generally applicable over the physical parameter ranges, independently of the coupling regimes. We may therefore interpret USC-DSC simply as the coupling regime where neither RWA nor CRWA applies and the general QRM dynamics must then be determined by the full QRM Hamiltonian $H_R\sim\frac{1}{2}(H_{JC}+\overline{H}_{AJC}~)$. In this respect, we may take advantage of the useful property that each component Hamiltonian $H_{JC}$ , $\overline{H}_{AJC}$ generates exact dynamical evolution to develop algebraic methods to disentangle the full QRM time evolution operator $U_{QRM}(t)\sim e^{-\frac{it}{2\hbar}(H_{JC}+\overline{H}_{AJC})}$ to determine the general dynamics generated by the full QRM Hamiltonian $H_R$.

\section{Acknowledgement}
I thank Maseno University for providing facilities and a conducive work environment during the preparation of the manuscript.

\end{document}